# On the Nature and Types of Anomalies: A Review of Deviations in Data

Ralph Foorthuis
Amsterdam, The Netherlands

*Abstract*: Anomalies are occurrences in a dataset that are in some way unusual and do not fit the general patterns. The concept of the anomaly is typically ill-defined and perceived as vague and domain-dependent. Moreover, despite some 250 years of publications on the topic, no comprehensive and concrete overviews of the different types of anomalies have hitherto been published. By means of an extensive literature review this study therefore offers the first theoretically principled and domain-independent typology of data anomalies and presents a full overview of anomaly types and subtypes. To concretely define the concept of the anomaly and its different manifestations, the typology employs five dimensions: data type, cardinality of relationship, anomaly level, data structure, and data distribution. These fundamental and data-centric dimensions naturally yield 3 broad groups, 9 basic types, and 63 subtypes of anomalies. The typology facilitates the evaluation of the functional capabilities of anomaly detection algorithms, contributes to explainable data science, and provides insights into relevant topics such as local versus global anomalies.

*Note*: Also see this video with animated 3D data plots illustrating the anomaly (sub)types: https://youtu.be/XOBR2sc00y4

## 1 Introduction

The physical and social world is known to bring about abnormal and bizarre phenomena that are seemingly hard to explain. Although rare by definition, such strange and unusual occurrences can actually also said to be relatively abundant due to the huge amount of objects and interactions in the world. Owing to the massive data collection taking place in the current era and the imperfect measurement systems used for this, anomalous observations can thus be expected to be amply present in our datasets. These large collections of data are mined in both academia and practice, with the aim of identifying patterns as well as peculiarities. The term *anomalies* in this context refers to cases, or groups of cases, that are in some way unusual and deviate from some notion of normality [1-13]. Such occurrences are often also referred to as outliers, novelties, deviants or discords [5, 14-16]. Anomalies are assumed to be both rare and different, and pertain to a wide variety of phenomena, which include static entities and time-related events, single (atomic) cases and grouped (aggregated) cases, as well as desired and undesired observations [7, 9, 16-21, 300, 319, 326]. Although anomalies can form a noise factor hindering the data analysis, they may also constitute the actual signals that one is looking for. Identifying them can be a difficult task due to the many shapes and sizes they come in, as illustrated in Fig. 1. *Anomaly detection* (AD) is the process of analyzing the data to identify these unusual occurrences. Outlier research has a long history and traditionally focused on techniques for rejecting or accommodating the extreme cases that hamper statistical inference. Bernoulli seems to be the first to address the issue in 1777 [22], with subsequent theory building throughout the 1800s [23-26, 327, 328], 1900s [27-36, 177, 274] and beyond [e.g. 37-39]. Although it was occasionally recognized that anomalies may be interesting in their own right [e.g. 12, 29, 33, 40-42], it was not until the end of the 1980s that they started to play a crucial role in the detection of system intrusions and other sorts of unwarranted behavior [43-50]. At the end of the 1990s another surge in AD research focused on general-purpose, non-parametric approaches for detecting interesting deviations [51-56]. Anomaly detection has now been studied for a wide variety of purposes, such as fraud discovery, data quality analysis, security scanning, system and process control, and – as indeed practiced in classical statistics for some 250 years – data handling prior to statistical inference [e.g. 3, 5, 14, 21, 24, 25, 57, 58, 158]. The topic of AD has not only gained ample academic attention over the years, but is also deemed crucial for industrial practice [59-63].

 ✉ ralph.foorthuis@heineken.com / ralphfoorthuis@gmail.com

Despite abundant research and valuable progress, the field of anomaly detection cannot claim maturity yet. It lacks an overall, integrative framework to understand the nature and different manifestations of its focal concept, the anomaly [6, 69, 184]. The general definitions of an anomaly are often said to be "vague" and dependent on the application domain [11, 12, 20, 64-68, 160, 316-318], which is likely due to the wide variety of ways anomalies manifest themselves. In addition, although the data mining, artificial intelligence and statistics literature does offer various ways to distinguish between different kinds of anomalies, research has hitherto not resulted in overviews and conceptualizations that are both comprehensive and concrete. Existing discussions on anomaly classes tend to be either only relevant for specific situations or so abstract that they neither provide a tangible understanding of anomalies nor facilitate the evaluation of AD algorithms (see sections 2.2 and 4). Moreover, not all conceptualizations focus on the intrinsic properties of the data and almost none of them use clear and explicit theoretical principles to differentiate between the acknowledged classes of anomalies (see section 2.2). Finally, the research on this topic is fragmented and studies on AD algorithms usually provide little insight into the kinds of anomalies the tested solutions can and cannot detect [6, 8, 184]. This literature study therefore presents an integrative and data-centric typology that defines the key dimensions of anomalies and provides a concrete description of the different types of deviations one may encounter in datasets. To the best of my knowledge this is the first comprehensive overview of the ways anomalies can manifest themselves, which, given that the field is about 250 years old, can be safely said to be overdue. The concept of the anomaly, including its different types and subtypes, is meaningfully characterized by five fundamental dimensions of anomalies, namely data type, cardinality of relationship, anomaly level, data structure, and data distribution. The value of the typology lies in offering a theoretical yet tangible *understanding* of the essence and types of data anomalies, assisting researchers with systematically *evaluating* and *clarifying* the functional capabilities of detection algorithms, and aiding in *analyzing* the conceptual characteristics and levels of data, patterns and anomalies. Preliminary versions of the typology have been employed for evaluating AD algorithms [6, 69, 70, 297]. This study extends the initial versions of the typology, discusses its theoretical properties in more depth and provides a full overview of the anomaly (sub)types it accommodates. Real-world examples from fields such as evolutionary biology, astronomy and – from my own research – organizational data management serve to illustrate the anomaly types and their relevance for both academia and industry.

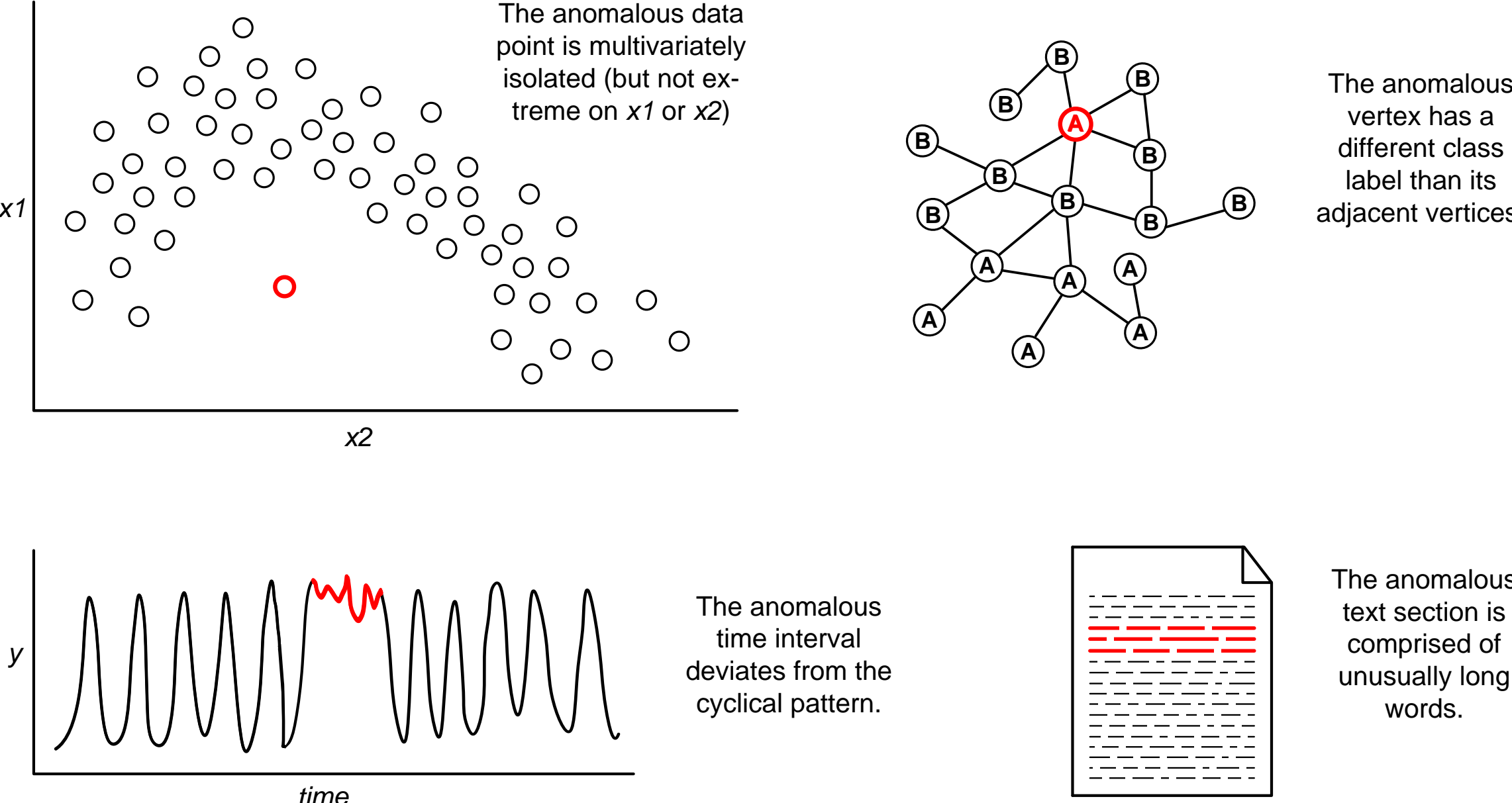

**Fig. 1: Red bold occurrences illustrate the wide variety of anomalies, resulting in the anomaly being perceived as an ambiguous concept. Resolving this requires typifying all these manifestations in a single overarching framework.**

A key property of the typology presented in this work is that it is fully data-centric. The anomaly types are defined in terms of characteristics intrinsic to data, thus without any reference to external factors such as measurement errors, unknown natural events, employed algorithms, domain knowledge or arbitrary analyst decisions. This is different from many other conceptualizations, as will be discussed in section 2.2 and 4. Note that "defining an anomaly type" in this context does not imply an ex ante domain-specific definition known before the actual analysis (e.g., based on rules or supervised learning). Unless specified otherwise, the anomalies discussed in this study can in principle be detected by unsupervised AD methods, thus based on the intrinsic properties of the data at hand, without any need for domain knowledge, rules, prior model training or specific distributional assumptions. Such anomalies are therefore universally deviant, regardless of the given problem.

A clear understanding of the nature and types of anomalies in data is crucial for various reasons. First, it is important in data mining, artificial intelligence and statistics to have a fundamental yet tangible understanding of anomalies, their defining characteristics and the various anomaly types that may be present in datasets. The typology's theoretical dimensions describe the nature of data and capture (deviations from) patterns therein, and as such offer a deep understanding of the field's focal concept, the anomaly. This is not only relevant for academia, but also for practical applications, especially now that AD has gained increased attention from industry [61-63]. Second, with the criticism on 'black box' and 'opaque' AI and data mining methods that may result in biased and unfair outcomes, it has become clear that it is often undesirable to have techniques and analysis results that lack transparency and cannot be explained meaningfully [71-76]. This is especially true for AD algorithms, as these may be used to identify and act on 'suspicious' cases [48-50, 326, 330]. Moreover, the definitions of anomalies are sometimes non-obvious and hidden in the designs of algorithms [8, 65, 184], and true deviations may be declared anomalous for the wrong reasons [306]. Although the typology presented here does not increase the transparency of the algorithms, a clear understanding of (the types of) anomalies and their properties, abstracted from detailed formulas and algorithms, does increase *post-hoc interpretability* by making the analysis results and data more understandable [20, 52, 69, 76, 184, 276]. Third, even if techniques from computer science and statistics are functionally transparent and understandable, the implementations of these algorithms may be done poorly or simply fail due to overly complex real-world settings [73, 77-79]. A clear view on anomalies is therefore needed to determine whether detected occurrences indeed constitute true deviations. This is especially relevant for unsupervised AD settings, as these do not involve pre-labeled data. Fourth, the *no free lunch* theorem, which posits that no single algorithm will demonstrate superior performance in all problem domains, also holds for anomaly detection [17, 60, 80-87, 184, 286, 320]. Individual AD algorithms are generally not able to detect all types of anomalies and do not perform equally well in different situations. The typology provides a functional evaluation framework that enables researchers to systematically analyze which algorithms are able to detect what types of anomalies to what degree. Fifth, a comprehensive overview of anomalies contributes to making implemented systems more robust and stable, as it allows injecting test datasets with deviations that represent unexpected and possibly faulty behavior [314, 329]. Finally, a principled overall framework, grounded in extant knowledge, offers students and researchers foundational knowledge of the field of anomaly analysis and detection, and allows them to position and scope their own academic endeavors.

This study therefore puts forward an overall typology of anomalies and provides an overview of known anomaly types and subtypes. Rather than presenting a mere summing-up, the different manifestations are discussed in terms of the theoretical dimensions that describe and explain their essence. The anomaly (sub)types are described in a qualitative fashion, using meaningful and explanatory textual descriptions. Formulas are not presented, as these often represent the detection techniques (which are not the focus of this study), and may draw attention away from the anomaly's cardinal properties. Also, each (sub)type can be detected by multiple techniques and formulas, and the aim is to abstract from those by typifying them on a somewhat higher level of meaning. A formal description would also bring with it the risk of unnecessarily excluding anomaly variations. As a final introductory remark it should be noted that, despite this study's extensive literature review, the long and rich history of anomaly research makes it impossible to include each and every relevant publication.

This article proceeds as follows. Section 2 explains key concepts and discusses related research. Section 3 introduces the typology of anomalies. Section 4 discusses various properties of the typology and compares it with other research. Finally, section 5 is for conclusions.

# 2 Theory

## 2.1 Key terms and concepts

This section defines the employed concepts to ensure that the reader understands the terms as intended, regardless of his or her discipline (senior scholars may choose to only do a quick scan). An anomaly, in its broadest meaning, is something that is different or peculiar given what is usual or expected [88-90]. In the philosophy of science, anomalies play a crucial role as observations or predictions that are inconsistent with the models in the prevailing academic paradigm [91-94]. Such anomalies require an explanation and consequently initiate the advancement of knowledge by the refinement of current theories. Over time, anomalies that constitute fundamental novelties may accumulate and trigger an academic crisis in which the old paradigm is replaced by a wholly different one. Newtonian physics, for example, was succeeded by Einstein's theory of general relativity, which was better capable of predicting and explaining a variety of observed astronomical phenomena, such as anomalies pertaining to the perihelion of Mercury. In statistics, data mining and AI an anomalous occurrence deviates from some notion of normality for the given data and setting. Deviants that can be detected in an unsupervised fashion, which are the focus of this study, can be defined more precisely. An *anomaly* in this context is a case, or a group of cases, that in some way is unusual and does not fit the general patterns exhibited by the majority of the data [3, 4, 8, 10, 11, 69, 325, 326]. The detection of anomalies is a highly relevant task, not only because they should be handled appropriately during inferential research, but also because the goal of analyses is often to discover interesting new phenomena [9, 37-39, 95-98]. The remainder of this section will focus on terms and concepts pertaining to anomalies in data.

The term *cases* refers to the individual instances in a dataset, also called data points, rows, records or observations [57, 99, 323]. These cases are described by one or more *attributes,* also referred to as variables, columns, fields, dimensions or features. Some of these attributes will be required for data management and context, such as identification (ID) and time variables. In addition, the dataset will contain substantive attributes, i.e. the meaningful domain-specific variables of interest, such as income and temperature. Measuring and recording the actual attribute values is prone to errors, the discovery of which may indeed be one of the reasons to conduct anomaly detection. The term *occurrence* is used here in a broad fashion and may refer to an individual case or a group of cases, an object or an event, and anomalous or regular data.

The term *dependency* is used in the literature to refer to two aspects of relationships, both of which are relevant for this study. First, there can be a dependency between the attributes, meaning there is a *relationship between the variables* [59, 96, 99, 100, 101, 182]. Income, for example, may be correlated with education and parental financial status. A second form of dependency, referred to as *dependent data*, deals with the relationship between the dataset's individual cases or rows [7, 20, 57, 102, 323]. A set with such dependent cases contains an intrinsic relation between the observations. Examples are time series, spatial and graph data, and sets with hierarchical relationships. The dependencies in such datasets are typically captured by time, location, linking or grouping attributes. These inter-case relations are absent from *independent data*, such as in i.i.d. random samples for cross-sectional surveys, in which every row represents a stand-alone observation.

Describing and understanding the different types of anomalies in a concrete and data-centric manner is not feasible without referring to the functional *data structures* that host them. This section therefore shortly discusses several important formats for organizing and storing data [cf. 5, 57, 95, 106, 110-115, 184]. Some analyses are conducted on *unstructured and semi-structured text documents*. However, most datasets have an explicitly structured format. *Cross-sectional data* consist of observations on unit instances – e.g. individual people, organizations or countries – at one point in time. The cases in such a set are generally considered to be unordered and otherwise independent, as opposed to the following structures with dependent data. *Time series data* consist of observations on one unit instance (e.g. one country) at different points in time. Time-oriented *panel data*, or longitudinal data, consist of a set of time series and are therefore comprised of observations on multiple individual entities at different points in time (e.g. income history for a sample of citizens followed over a five-year period). The general term *sequence data* will be used to refer to time series, time-oriented panel data, as well as to sets with an ordering not based on time. Sequence data have broad applications and, besides time-oriented phenomena, are able to capture genomic and other biological features, user actions, spectroscopy wavelengths, trajectories, audio, and even visual information such as the shape of physical objects and moving elements in a video [5, 10, 114, 116-123, 278]. Each of the above data structures can be implemented with a single matrix or table, but when several inter-related entities need to be modelled a *relational model* is often used [95, 124, 125]. This allows many functional structures, including domain-specific designs and analytical star schemas [95, 103, 104]. A *graph* is a related data structure and typically consists of vertices (nodes), edges (connections or links), edge directions and edge weights [20, 57, 95, 106, 112, 113]. An attributed graph has, in addition to these structural properties, any number of substantive domain-specific variables. Such structures are highly relevant for modelling e.g. social networks, chemical compounds, Internet data and wireless sensor networks. Graphs can take many forms, including *tree* data structures. A tree consists of a root, a given amount of parent and child nodes, and does not feature any closed paths (so-called cycles). Storing graphs typically involves both a set of vertices (e.g. as a list of nodes and their properties) and a set of edges (e.g. as an adjacency matrix with relations, directions and weights). This is similar to *spatial data*, which usually consist of a set of coordinates and a set of substantive features [8, 66, 95, 126-129, 277, 323]. The latter may pertain to e.g. population density, settlement type or availability of utility infrastructure elements. The representations often constitute points (atomic positions such as addresses), lines, arcs (e.g. roads or rivers) and polygons (regions such as neighborhoods or states). However, a rasterization of continuous data can also be used, such as satellite imagery and brain scans represented as a grid of pixels. This format therefore also captures image material in general, with the data points being 2D pixels or 3D voxels. The coordinates represent a position on a canvas or frame, while the features store the visual information as gray intensities or color information (e.g. RGB, multi- or hyperspectral). *Spatio-temporal data* feature a sequence dimension in addition to the coordinates, and may capture e.g. video and historical geographical information [281, 283]. Even more detailed distinctions can be made, but the key formats described above suffice to properly discuss the wide variety of anomaly types.

The concept of an *aggregate* is often used in the context of dealing with noise or obtaining a more abstract representation at the level of interest. When aggregating *the cases* the analyst is able to treat multiple individual rows as a whole or a group, and consequently obtain summary statistics – e.g. means and totals – or other derived properties of the collective [5, 57, 95, 103-106, 282, 323]. This allows the data and perspective to be transformed, for example, from days to months or from individual people to households. In terms of data structures, typical examples of aggregates are subgraphs, subsequences and regions in spatial data. Often, however, the AD analysis will simply focus on the individual cases, i.e. on the atomic level of the set's microdata. One can also aggregate *the attributes* in order to reduce dimensionality or to obtain for individual or grouped cases meaningful and complex latent variables or manifolds [3, 96, 107-109]. However, this may mainly be relevant in the context of high-level semantic anomalies, in which such aggregates are generally difficult to determine without any prior theory, rules or supervised training (see section 3.2 for more information) [13, 310].

To conclude this section it is valuable to briefly discuss what constitutes a *typology*. To theoretically distinguish between concepts, scholars have various intellectual tools at their disposal, amongst which are taxonomies, classifications, dendrograms and typologies [130]. These all make use of one or more *classificatory principles* (explicit dimensions) to differentiate between the relevant elements. A classification uses a single principle, whereas a typology uses two or more simultaneously. A typology is therefore well-suited to theoretically distinguish between complex concept types – offering not only a fundamental and summarized description of a general concept, but also an exhaustive and mutually exclusive overview of its distinct but related types. The term *classification* will be used more loosely in this study and will also refer to conceptualizations with classes that are not based on clear principles and that are neither mutually exclusive nor jointly exhaustive.

## 2.2 Related work

The literature acknowledges various ways to distinguish between different manifestations of anomalies. Barnett and Lewis [2, cf. 31, 131] make a distinction between *extreme but genuine members of the main population*, i.e. random fluctuations at the tails of the focal distribution, and *contaminants*, which are observations from a different distribution. Wainer [34] differentiates between *distant outliers*, which exhibit extreme values and are clearly in error, and *fringeliers*, which are unusual but with their position about three standard deviations from the majority of the data cannot be said to be extremely rare and unequivocally erroneous. Essentially the same distinction is made in [132] with *white crows* and *in-disguise anomalies* respectively. Relatedly, in [5, 133] a distinction is made between a *weak outlier* (noise) and a *strong outlier* (a significant deviation from normal behavior). The latter category can be sub-divided in *events,* i.e. unusual changes in the real-world state, and *measurement errors,* such as a faulty sensor [134, 135]. An overall classification is presented in [96], with the classes of anomalies indicating the underlying reasons for their deviant nature: a *procedural error* (e.g. a coding mistake), an *extraordinary event* (such as a hurricane), an *extraordinary observation* (unexplained deviation), and a *unique value combination* (which has normal values for its individual attributes). Other sources refer to similar explanations in a more free-format fashion [39, 97, 136]. In [184] a distinction is made between 9 types of anomalies. Another broad classification is that of [7], which differentiates between three general categories. A *point anomaly* refers to one or several individual cases that are deviant with respect to the rest of the data. A *contextual anomaly* appears normal at first, but is deviant when an explicitly selected context is taken into account [cf. 137].

| Reference | G/S | DC? | Classes of anomalies | Explicit classificatory dimensions |
|---|---|---|---|---|
| [6, 69, 70] | G | Y | Extreme value ano, Rare class ano, Simple mixed data ano, Multidimensional numerical ano, Multidimensional rare class ano, Multidimensional mixed data ano | Types of data, Cardinality of relationship |
| [2, cf. 31] | G | N | Extreme genuine member, Contaminant | None |
| [34] | G | Y | Fringelier, Distant outlier | None |
| [52] | G | Y | Strongest outlier, Weak outlier, Trivial outlier | Attribute subspace |
| [132] | G | Y | White crow, In-disguise anomaly | None |
| [5, 133] | G | Y | Weak outlier, Strong outlier | None |
| [96] | G | N | Procedural error, Extraordinary event, Extraordinary observation, Unique value combination | None |
| [136] | G | N | Data error, Normal variance, Data from other distributions, Distributional assumption | None |
| [7] | G | N | Point anomaly, Contextual anomaly, Collective anomaly | None |
| [184] | G | Y | Known distribution ano, Sparse distribution ano, Local density-based ano, Global density-based ano, Rare instance ano, Burst ano, Deviant sequence ano, Trend ano, Irregularity ano | None |
| [182] | G | Y | Trivial outlier, Non-trivial outlier | None |
| [3, 143] | S | N | Outlier, High-leverage point, Influential point | None |
| [138, cf. 141] | S | Y | Additive outlier, Temporary change, Level shift, Innovational outlier | None |
| [187] | S | Y | Isolated outlier, Patch outlier, Level shift | None |
| [142] | S | Y | Isolated outlier, Shift outlier, Amplitude outlier, Shape outlier | None |
| [233] | S | Y | Trend anomaly, Seasonality anomaly | None |
| [314] | S | Y/N | Outlier, Spike, Stuck-at, High-noise (plus several non-data-centric anomalies) | None |
| [281] | S | Y | Various spatio-temporal change patterns | Temporal, Spatial, Raster/Vector |
| [20] | S | Y | Deviant vertex, Deviant edge, Deviant subgraph | None |
| [205] | S | Y | Near-star, Near-clique, Heavy vicinity, Dominant edge | None |
| [125] | S | Y | Insertion, Update and Deletion anomaly | Based on database CRUD functions |
| [60] | S | Y | Foreign-symbol, Foreign n-gram, Rare n-gram | None |
| [118, cf. 146] | S | Y | Positional outlier, Angular outlier | None |

**G/S** refers to a general (broad and usually abstract) versus specific way to distinguish between classes of anomalies. **DC** stands for data-centric, meaning the anomalies can be distinguished by analyzing the dataset, without a reference to or dependency on external factors (such as unknown real-world events or arbitrary analyst decisions).

**Table 1. Existing classifications for distinguishing between anomalies.**

An example is a temperature value that is only remarkably low in the context of the summer season. Finally, a *collective anomaly* refers to a collection of data points that belong together and, as a group, deviate from the rest of the data.

Several specific and concrete classifications are also known, especially those dedicated to sequence and graph analysis. Many of their anomaly types will be described in detail in section 3. In time series analysis several within-sequence types are acknowledged, such as the *additive outlier*, *temporary change*, *level shift* and *innovational outlier* [138-141, 191]. The taxonomy presented in [142] focuses on between-sequence anomalies in panel data and makes a distinction between *isolated outliers*, *shift outliers, amplitude outliers* and *shape outliers*. Another specific classification is known from regression analysis, in which it is common to distinguish between *outliers*, *high-leverage points* and *influential points* [3, 143-145]. Two anomalies with regard to the trajectories of moving entities are presented in [118, cf. 146, 147], namely the *positional outlier*, which is positioned in a low-density area of the trajectory space, and the *angular outlier*, which has a direction different from regular trajectories. The subfield of graph mining has also acknowledged several specific classes of anomalies, with anomalous *vertices*, *edges* and *subgraphs* being the basic forms [18, 20, 112, 113, 148, 149]. Table 1 summarizes the anomaly classes acknowledged in the extant literature. In section 3 these anomalies, particularly those that allow a data-centric definition, will be discussed in more detail and positioned within this study's typology.

The classifications in Table 1 are either too general and abstract to provide a clear and concrete understanding of anomaly types, or feature well-defined types that are only relevant for a specific purpose (such as time series analysis, graph mining or regression modeling). The fifth column also makes clear that extant overviews hardly offer clear principles to systematically partition the classificatory space to obtain meaningful categories of anomalies. They thus do not constitute a classification or typology as defined by [130]. To the best of my knowledge this study's framework and its predecessors offer the first overall typology of anomalies that presents a comprehensive overview of concrete anomaly types.

Many of the existing overviews also do not offer a data-centric conceptualization. Classifications often involve algorithm- or formula-dependent definitions of anomalies [cf. 8, 11, 17, 86, 150, 184], choices made by the data analyst regarding the contextuality of attributes [e.g. 7, 137], or assumptions, oracle knowledge, and references to unknown populations, distributions, errors and phenomena [e.g. 1, 2, 39, 96, 131, 136]. This does not mean these conceptualizations are not valuable. On the contrary, they often provide important insights as to the underlying reasons why anomalies exist and the options that a data analyst can exploit. However, this study exclusively uses the intrinsic properties of the data to define and distinguish between the different sorts of anomalies, because this yields a typology that is generally and objectively applicable. Referencing external and unknown phenomena in this context would be problematic because the true underlying causes usually cannot be ascertained, which means distinguishing between e.g. extreme genuine observations and contaminants is difficult at best and subjective judgments necessarily play a major role [2, 4, 5, 34, 314, 323]. A data-centric typology also allows for an integrative and all-encompassing framework, as all anomalies are ultimately represented as part of a data structure. This study's principled and data-oriented typology therefore offers an overview of anomaly types that not only is general and comprehensive, but also comes with tangible, meaningful and practically useful descriptions.

To end this section it is good to note that many valuable classifications of anomaly detection *techniques* are available [5, 7, 13, 14, 55, 84, 135, 150-152, 299-301, 318-320, 330]. Because the core focus of the current study is on *anomalies*, detection techniques are only discussed if valuable in the context of the typification of data deviations. A review of AD techniques is therefore out of scope, but note that the many references direct the reader to information on this topic.

# 3 A Typology of Anomalies

## 3.1 Classificatory principles

This section presents the five fundamental data-oriented dimensions employed to describe the types and subtypes of anomalies: data type, cardinality of relationship, anomaly level, data structure, and data distribution. The typology's framework, as depicted in Fig. 2, comprises three main dimensions, namely data type, cardinality of relationship and anomaly level, each of which represents a classificatory principle that describes a key characteristic of the nature of data [57, 96, 100, 106]. Together these dimensions distinguish between nine basic anomaly types. The first dimension represents the **types of data** involved in describing the behavior of the occurrences. This pertains to these data types of the attributes responsible for the deviant character of a given anomaly type [10, 57, 96, 97, 114, 161]:

- *Quantitative:* The variables that capture the anomalous behavior all take on numerical values. Such attributes indicate both the possession of a certain property and the degree to which the case may be characterized by it, and are measured at the interval or ratio scale. This kind of data generally allows meaningful arithmetic operations, such as addition, subtraction, multiplication, division and differentiation. Examples of such variables are temperature, age and height, which are all continuous. Quantitative attributes can also be discrete, however, such as the amount of people in a household.
- *Qualitative*: The variables that capture the anomalous behavior are all categorical in nature and thus take on values in distinct classes (codes or categories). Qualitative data indicate the presence of a property, but not the amount or degree. Examples of such variables are gender, country, color and animal species. Words in a social media stream and other symbolic information also constitute qualitative data. Identification attributes, such as unique names and ID numbers, are categorical in nature as well because they are essentially nominal (even if they are technically stored as numbers). Note that although qualitative attributes always have discrete values, there can be a meaningful order present, such as with the ordinal martial arts classes 'LIGHTWEIGHT', 'MIDDLEWEIGHT' and 'HEAVYWEIGHT'. However, arithmetic operations such as subtraction and multiplication are not allowed for qualitative data.
- *Mixed*: The variables that capture the anomalous behavior are both quantitative and qualitative in nature. At least one attribute of each type is thus present in the set describing the anomaly type. An example is an anomaly that involves both country of birth and body length.

| | | **Types of Data** | | | | |
|---|---|---|---|---|---|---|
| | | **Quantitative attributes** | **Qualitative attributes** | **Mixed attributes** | | |
| **Cardinality of Relationship** | **Univariate** | Type I<br>Uncommon number anomaly | Type II<br>Uncommon class anomaly | Type III<br>Simple mixed data anomaly | **Atomic** | **Anomaly Level** |
| | | Atomic univariate anomaly | | | | |
| | **Multivariate** | Type IV<br>Multidimensional numerical anomaly | Type V<br>Multidimensional categorical anomaly | Type VI<br>Multidimensional mixed data anomaly | | |
| | | Atomic multivariate anomaly | | | | |
| | | Type VII<br>Aggregate numerical anomaly | Type VIII<br>Aggregate categorical anomaly | Type IX<br>Aggregate mixed data anomaly | **Aggregate** | |
| | | Aggregate anomaly | | | | |

**Fig. 2: The framework for the typology of anomalies**

The second dimension is the **cardinality of relationship**, which represents how the various attributes relate to each other when describing anomalous behavior. These attributes are *individually* or *jointly* responsible for the deviant character of the occurrences [39, 59, 96, 100, 105, 106, 136, 158, 285]:

- *Univariate*: Except for being part of the same set, no relationship between the variables exists to which the anomalous behavior of the deviant case can be attributed. To describe and detect the anomaly, its variables can therefore be referred to separately. In other words, the analysis can assume independence between the attributes.

- *Multivariate*: The deviant behavior of the anomaly can be attributed to the relationship between its variables. The anomaly needs to be described and detected by referring to the joint distribution, meaning the individual attributes cannot be studied separately. Variables have to be analyzed jointly in order to take into account their relation, i.e. their combination of values. The term 'relationship' should be interpreted broadly here and includes correlations, partial correlations, interactions, collinearity, concurvity, (non)collapsibility, and associations between attributes of different data types.

The cardinality of relationship essentially refers to whether one attribute is sufficient to define and detect the anomaly type or that multiple attributes need to be taken into account simultaneously. Note that, owing to the massive data collection nowadays, a dataset is likely to contain many attributes beyond the *hosting subspace* (i.e. the subset of attributes required to describe and detect a given anomaly). As a matter of fact, an occurrence can be deviant in one subspace and normal in others [133, 162-164, 180, 182, 303]. An occurrence could even be one type of anomaly in one subspace and another type in a second subspace.

The third dimension is the **anomaly level**, which represents the distinction between *atomic* anomalies (individual low-level cases or data points) versus *aggregate* anomalies (groups or collective structures). In theory this is also an independent dimension, but in practice univariate data only contain atomic anomalies. Multivariate data may also host aggregate anomalies, which alongside substantive attributes typically require data management attributes (e.g. time stamps or group designations) that allow the formation of collective structures. However, should future research introduce noteworthy examples of aggregate anomalies in univariate data, the framework can be easily extended to accommodate this.

The fourth dimension represents the **data structure**, which is used to distinguish between the subtypes within the nine cells of the typology. A given cell may contain multiple anomaly subtypes, which have defining characteristics that can be traced back to the specific data formats that host them, such as graphs and time series. Also note that the difference between dependent and independent data is treated as a characteristic of the data structure here. See section 2.1 for an overview of the different structures.

The fifth dimension is the **data distribution**, which refers to the collection of attribute values and their pattern or dispersion throughout the data space [98, 165]. An anomaly, per definition, is defined by its difference with regard to the remainder of the data, which makes the distribution of the dataset an important factor to take into account. The distribution is strongly dependent on the classificatory factors mentioned above, but allows focusing on density and other dispersion-related aspects of the set. It therefore offers additional descriptive and delineating capabilities.[1] This dimension will not only be used to subdivide between anomaly subtypes within the typology's nine cells, but occasionally also to illustrate how an altered distribution would result in a different manifestation of a given anomaly.

The five classificatory principles of the typology are not only fundamental in the sense that they describe theoretically crucial properties of data, but also because they deeply impact analysis and storage solutions. Some examples of this: jointly analyzing qualitative and quantitative data requires specialized multivariate techniques; analyzing dependent data usually needs to account for autocorrelation; and locating clusters and other patterns in multidimensional data implies discovering inter-variable relationships and dealing with exponential scaling issues as datasets increase in size.

The preliminary typology presented in [69, 70, cf. 6] is summarized in the first row of Table 1 (it essentially lacked the lowest layer for aggregate anomalies present in Fig. 2). Although this version was able to implicitly accommodate complex anomalies, several discussions at conferences pointed to the fact that the types in the multivariate row were rather broad and demanded further subdivision and clarification. Atomic and aggregate anomalies are therefore acknowledged explicitly in the framework now, yielding nine basic anomaly types. In addition, some of the terminology is updated. The new typology framework is depicted in Fig. 2. Detailed subtypes are included in Fig. 3 and illustrated in the data plots throughout this article. The nine main types of anomalies, which follow naturally and objectively from the classificatory principles, are described in section 3.2. (Note on visualization: in many of the diagrams the data points' colors and shapes represent different categorical values; the reader might want to zoom in on a digital screen to see colors, shapes and patterns in detail.)

[1] The term "distribution" is usually not explicitly defined. One could argue that the concept of data distribution alone is sufficient to describe anomalies. However, such a simplified stance would defeat the purpose of this study, namely to offer fundamental and concrete insights into the nature and types of anomalies. The distribution here thus excludes the other dimensions.

## 3.2 Overview of anomaly types and subtypes

This section presents the anomaly types and their concrete subtypes. The typology's rows represent three broad groups of anomaly types. *Atomic univariate anomalies* are single cases with a deviant value for one or possibly multiple individual attributes. They are relatively easy to describe and detect because the individual values of these observations are unusual. Relationships between attributes or cases are not relevant for such occurrences. An example is an extremely high numerical value, such as a person reported to be 246 cm high. There may be several anomalous atomic occurrences (albeit apparently not in a way so as to form a 'normal pattern'), but the essence is that each individual case is anomalous in its own right. Moreover, should an individual case have multiple unusual values, then each of them will be anomalous (e.g. not only 246 cm high, but also 117 years old). *Atomic multivariate anomalies* are single cases whose deviant nature lies in their relationships, with the individual values not being anomalous. In independent data this will manifest itself in the unusual combination of a case's own attribute values, such as a 10 year old person with a body length of 180 cm. However, the multivariate nature also allows defining and detecting deviations in dependent data, i.e. in the relation with the other cases to which the given case is linked. An example is a time series temperature measurement that is unusually high for winter, but that would have been normal in summer. Atomic multivariate anomalies hide in multidimensionality, as they cannot be described and detected by simply analyzing the individual variables separately. Finally, *aggregate anomalies* are groups of cases that deviate as a collective, of which the constituent cases usually are not individually anomalous. Relationships between attributes and between cases play a key role here, not only to position an occurrence in the set with dependent data, but often also to form a pre-defined or derived group. Owing to their complex and intricate nature, these anomalies are generally the most difficult to describe and detect. A deviant subsequence is one manifestation of an aggregate anomaly, such as a whole winter with many unusually high temperatures compared to other winters. Although the above may be abstract at first reading, the detailed discussion below will offer a concrete understanding.

Before presenting the individual types and subtypes it is worthwhile to make some assumptions explicit. It is assumed that an atomic case (individual row) with $p$ attributes represents *a single data point* in $p$-dimensional space (not a set of $p$ distinct data points). A time series consists of *multiple atomic data points*, each of which has $p$ attributes. The typology also assumes a *parsimonious data structure*, without redundant information. For example, the degree of a vertex, i.e. the number of edges that connect it to other vertices in the graph, is a latent characteristic of which the value is seen as being derived runtime during the analysis and is therefore assumed not to be explicitly included in the original dataset. When relevant the set does *explicitly include management and dependency attributes*, such as ID, sequence, group and link information, as these represent crucial structural properties (note that in the special case of text or symbolic data the sequence may be present implicitly). The typology also assumes the unaltered and *original dataset* in which one aims to declare anomalies. The reason for this is that e.g. normalization [16, 86], dimensionality reduction [166], log transformations [167] and data type conversions [70] have all been shown to have significant impact on the presence and detection of anomalies. To be sure, transformations are allowed, but the typology then either assumes the newly derived dataset as the starting point for typification or remains agnostic as to any transformations performed as part of the AD algorithm. Finally, if one needs to choose between potential anomaly types, then the norm is to opt for *the simplest type* that captures the deviant occurrence (see the Discussion for more on this).

The types and subtypes are visualized schematically in Fig. 3 and discussed in detail in the remainder of this section. Since even the subtypes can be quite broad when multivariate in nature, ample examples are also provided.

Types of Data

Legend
Normal point or object
Anomalous point or object
Independent data
Dependent data

Quantitative attributes
Qualitative attributes
Mixed attributes

Cardinality of Relationship
Univariate
Multivariate

Anomaly Level
Atomic
Aggregate

Type I: Uncommon number anomaly
a) Extreme tail value
b) Isolated intermediate value

Type II: Uncommon class anomaly
a) Unusual class
b) Deviant repeater

Type III: Simple mixed data anomaly
a) Extreme tail uncommon class
b) Intermediate uncommon class

Atomic univariate anomaly

Type IV: Multidimensional numerical anomaly
a) Peripheral point
b) Enclosed point
c) Local density anomaly ←
d) Global density anomaly →
e) Local additive anomaly
f) Deviant numerical spatial point (typically in images)
g) Deviant numerical spatio-temporal point (typically in videos)

Type V: Multidimensional categorical anomaly
a) Uncommon class combination
b) Deviant categorical vertex
c) Deviant categorical edge

Type VI: Multidimensional mixed data anomaly
a) Incongruous common class
b) Incongruous common sequential class
c) Deviant vertex
d-f) Unusual vertex insertion/change/removal
g) Deviant edge
h-j) Unusual edge insertion/change/removal
k) Deviant spatial point (typically in geo data)
l) Deviant spatio-temporal point (typically in geo data)

Atomic multivariate anomaly

Type VII: Aggregate numerical anomaly
a) Deviant cycle
b) Temporary change
c) Level shift
d) Innovational outlier
e) Trend change →
f) Variation change ←
g) Deviant numerical spatial region (typically in images)
h) Deviant numerical spatio-temporal region (typically in videos)
i) Point-based aggregate anomaly ←
j) Distribution-based aggregate anomaly →

Type VIII: Aggregate categorical anomaly
a) Deviant class aggregate (typically in texts)
b) Deviant categorical subgraph
c) Deviant relational aggregate

Type IX: Aggregate mixed data anomaly
a) Class change
b) Deviant class cycle
c) Deviant class sequence
d-i) Deviant isolation/shift/shape/amplitude/trend/variation sequence
j) Deviant subgraph
k-r) Appearing/disappearing/flickering/merging/splitting/growing/shrinking/eccentric (sub)graph
s) Deviant spatial region (typically in geo data)
t) Deviant spatio-temporal region (typically in geo data)
u) Point-based mixed data aggregate anomaly ←
v) Distribution-based mixed data aggregate anomaly →

Aggregate anomaly

**Fig. 3: The typology including all types and subtypes. Each anomaly subtype is represented by an icon that depicts the essence of the deviation. An icon that includes lines represents a set with dependent data. (Zoom in on a digital screen to see details.)**

### 3.2.1 Atomic univariate anomalies

This section provides an overview of anomaly types that consist of a single case with a deviant value for one or possibly several attributes, with each individual value being deviant in its own right. The more unusual a value is or the more attributes take on unusual values, the more anomalous the respective case is.

**I. Uncommon number anomaly**: This is a case with an extremely high, low or otherwise unusual value for one or multiple individual quantitative attributes [5, 97, 168]. These deviant numbers often manifest themselves as an *extreme tail value* (depicted as subtype ST-Ia in Fig. 3). They are hosted by the given attribute's numerical vector, which may contain one or more extreme values at the far ends of its statistical distribution. Fig. 4 shows two plots of the national Polis administration of Dutch income transactions [6], with various ST-Ia occurrences. Fig. 5 and Fig. 6 present examples as well. Traditional univariate statistics typically offers methods to detect this subtype, e.g. by using a measure of central tendency and a given degree of variation [5, 7, 30, 96, 97, 143, 153, 184]. Cases that clearly exceed a threshold are considered extreme and very distant ST-Ia instances. It follows that cases lying near this decision boundary, so-called 'fringeliers', are more difficult to interpret [34, cf. 5, 132]. Extreme tail values are literally 'outliers', as they lie in an isolated region of the numerical space. However, this does not necessarily mean the case is located far from the other data. An example of an extreme value lying relatively near the bulk of the data points can be found in real-world income data, as depicted at the left of Fig. 5 (which in this context may point to an error or improvised corrective transaction). Moreover, given a different data distribution, isolated low-density values can also be located in the middle of the value range [5, 59, 169, 307]. These *isolated intermediate values* (subtype ST-Ib) do not only lie outside the dense regions, but also in between them. They can manifest themselves, for example, in multimodal or disjoint probability distributions, where they may be extreme members of one of the populations. Traditional AD techniques for ST-Ia anomalies often cannot detect ST-Ib cases (see the Grubbs and GESD example in the Discussion).

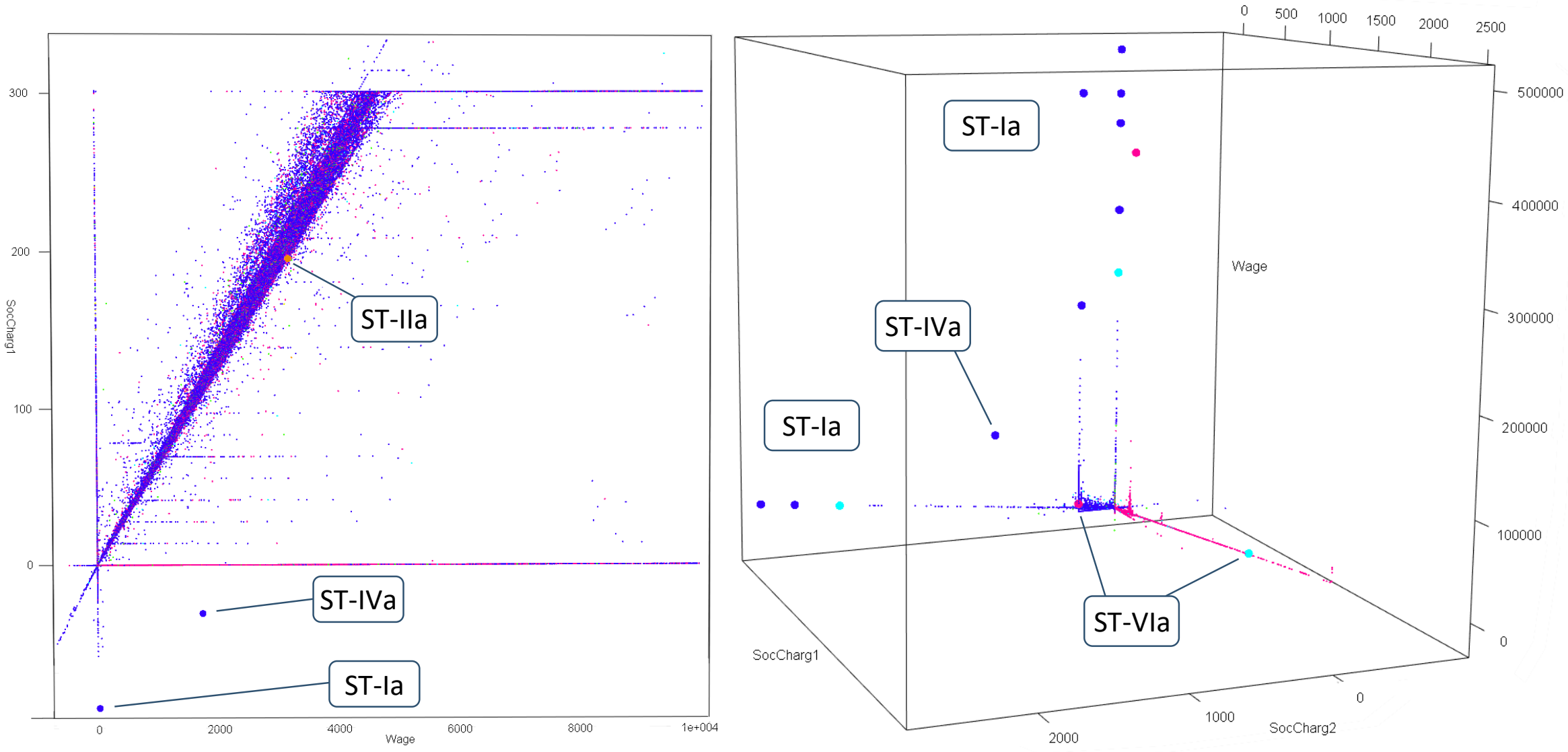

**Fig. 4: Real-world income data from the Polis administration with anomalies shown as large dots. The left plot has two and the right plot three numerical variables (wage and social charges). The social security code attribute is represented by color.**

The distribution of the variable affects the way Type I anomalies can manifest themselves in other ways as well [1, 153, 170, 171]. Skewed distributions [39, 167, 172], leptokurtic distributions [173, 174] and heavy-tailed distributions [131, 175, 176] tend to generate substantially more extreme cases than normal distributions do. Masking and swamping are also relevant from a distributional perspective [1, 2, 28, 36, 177, 178, 179].

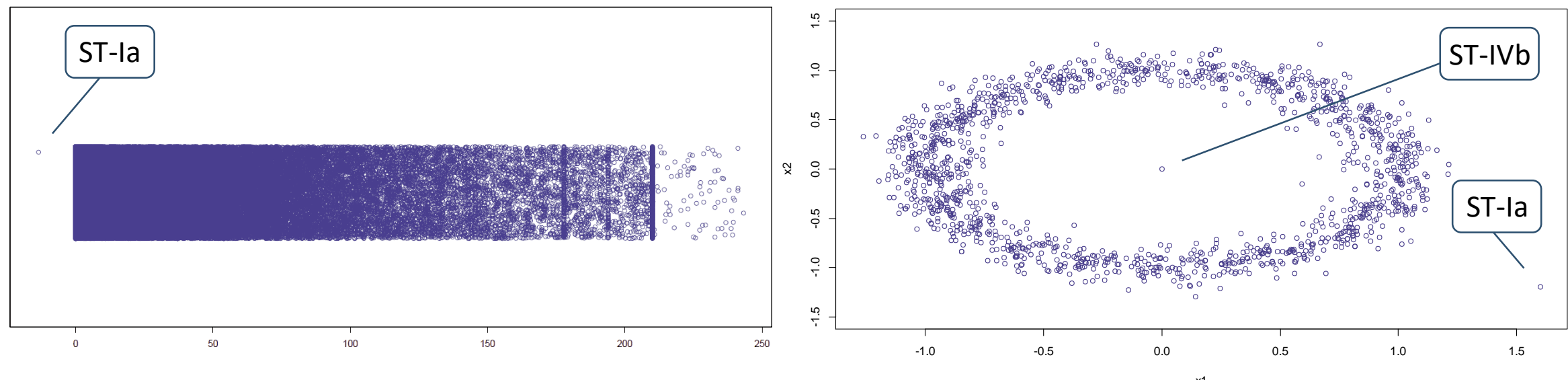

**Fig. 5: (Left) Univariate social charges data from the Polis administration. Note that the vertical dimension represents random scatterplot jitter for visualization purposes. (Right) Two-dimensional synthetic dataset.**

**II. Uncommon class anomaly**: The *unusual class* (ST-IIa) is a case with a unique or rare categorical value for one or several individual qualitative variables. The studies [60, 184, 309] discuss this type of anomaly. Case *ST-IIa* in Fig. 4 is a truly unique class, with the orange color representing the sole instance of the respective categorical value. The two *ST-IIa* cases in the left panel of Fig. 6 are the only square and reversed triangle. The red and orange colors of the *ST-IIa* points in the right panel of Fig. 6 are non-unique rare code values. These may not be clear-cut anomalies and thus may at some moment in the AD process demand that the concept of rarity is arbitrarily defined, e.g. by using a threshold [60, 85, cf. 34]. The *deviant repeater* (ST-IIb) is category that occurs frequently while the norm is to be non-frequent due to a highly skewed distribution. Such anomalies can occur in e.g. identification, IP address or name attributes [181].

**III. Simple mixed data anomaly**: This is a case that is both a Type I and a Type II anomaly, i.e. with at least one isolated numerical value and one uncommon class. The subtype *extreme tail uncommon class* (ST-IIIa) has a rare or unique class value at the tail of the distribution, whereas the subtype *intermediate uncommon class* (ST-IIIb) has an unusual class at an isolated intermediate location in the numerical space. Case *ST-IIIa* in Fig. 6 is an example. A Type III anomaly deviates with regard to multiple data types, requiring deviant values for at least two attributes, each anomalous in its own right [69, 70]. However, like Type I and II anomalies, analyzing the attributes jointly is unnecessary because the case in question is not multivariately anomalous. In other words, this type requires a set of individually deviant attribute values, not a deviant combination of attribute values. This is fundamentally different from the types described in 3.2.2.

### 3.2.2 Atomic multivariate anomalies

This section discusses anomalies that comprise a single case with a deviant combination of attribute values. In dependent data the deviancy typically lies in the relationship between the cases.

**IV. Multidimensional numerical anomaly**: This is a case that does not fit the general patterns when the relationship between multiple quantitative attributes is taken into account, without showing unusual values for any of the individual attributes that partake in this relationship. Type IV anomalies may reside not only in independent data, but also in dependent data because the multivariate character of the set allows taking into account the inter-case relationships. In independent data the anomalous nature of a case of this type lies in the unusual combination of its numerical attribute values [38, 39, 52, 182, 185]. Several quantitative attributes therefore need to be jointly taken into account to describe and detect such an anomaly. An example is a person who is 182 cm tall and weighs 53 kilos, i.e. an unusual combination of normal individual values [34]. In independent data such a case is literally ‘outlying’ from the relatively dense multivariate clouds and is thus located in an isolated area [cf. 101, 179]. This can be a *peripheral point* (ST-IVa), such as illustrated by the *ST-IVa* cases in Fig. 4 and Fig. 6. A second subtype is the *enclosed point* (ST-IVb), which means the anomaly is surrounded by normal data. An example is an anomaly located inside an annular region [302, 303, 307], illustrated by case *ST-IVb* in Fig. 5. Another example is a case in a spiral shape or other maze-like distribution [305]. Some methods, such as one-class support vector machines or iForest, are able to detect peripheral points, but are not geared towards identifying enclosed points [6, 305].

Another question becomes relevant if the dataset contains multiple clusters that have different densities. The outlyingness of an individual case can then be seen as being dependent upon the degree of isolation relative to its local area rather than to the global space [8, 17, 53, 55]. A *local density anomaly* (ST-IVc) is a case that is only isolated in the context of its neighborhood. Techniques to detect these anomalies, such as LOF and LOCI, need to account for the density of both the case in question and its neighbors. See the Discussion for more on the topic of locality. A dataset may also predominantly consist of random noise or large clusters, except for a few data points located close to each other. These points are *global density anomalies* (ST-IVd). Although they could be perceived as a single (albeit tiny) cluster, they are often conceptualized and detected as individual cases [9, 184, 305, 307]. ST-IVc and ST-IVd occurrences could in principle also have univariate equivalents, although these do not seem to be discussed in the literature.

For independent data the description of a Type IV instance requires multiple substantive attributes, as illustrated in the examples provided above. With dependent data the anomaly may also be defined by a single substantive variable, e.g. temperature, although at least one other attribute is typically still needed to link the related individual cases. In a quantitative context this usually concerns time series data, comprising a variable for ordered linking and a numerical substantive measure such as weight, wage, volume or heart rate [16, 114]. The *local additive anomaly* (ST-IVe) captures anomalous observations in time series and other numerical sequences. It features a short-lived spike that deviates from the local temporal neighborhood – e.g. the current season or trend – without exhibiting globally extreme values [cf. 141, 187-189]. As such this subtype implies that the substantive and sequence attributes are acknowledged and described jointly, i.e. multivariately. Case *ST-IVe* in Fig. 8.A is an example. Note that a globally extreme occurrence, such as case *ST-Ia* in the same figure, is simply a Type I extreme tail value for which the time attribute is irrelevant (unless the position in the sequence should be known). If the anomalous event transpires slowly relative to the measurement resolution, it may span multiple observations and should be considered an aggregate Type VII occurrence. However, the classic definition of an additive anomaly is that it is an abrupt change that pertains to a single observation, i.e. a Type IV anomaly [21, 105, 138, 141, 193].

A *deviant numerical spatial point* (ST-IVf) is a case that is unusual due to its quantitative spatial and possibly substantive features. If time is also a relevant factor the case is a *deviant numerical spatio-temporal point* (ST-IVg). Due to their quantitative nature these anomalies typically reside in images and videos respectively [68, 122, 194]. An anomaly then is an individual pixel or voxel that, given its location in the frame and possibly in time, has an unusual color or multispectral measurement. Anomalies are known to occur at this granular data point level, for example in satellite imagery [67, 68]. However, anomalies in this context are usually aggregates (e.g. a group of pixels), so this topic will be discussed in detail in the section on Type VII deviations. ST-IVf and ST-IVg occurrences can also be geographical anomalies, but examples thereof will be discussed as Type VI cases because they usually reside in mixed data.

**V. Multidimensional categorical anomaly**: This is a case that does not fit the general patterns when the relationship between multiple qualitative attributes is taken into account, without showing unusual values for any of the individual attributes that partake in this relationship. In short, a case with a rare or unique combination of class values, which can reside in independent or dependent data. In independent data two or more substantive categorical attributes from the same case need to be jointly taken into account to describe and identify a multidimensional categorical anomaly. An example is this curious combination of values from three attributes used to describe dogs: ‘MALE’, ‘PUPPY’ and ‘PREGNANT’. A visual example is case *ST-Va* in Fig. 6, as it is the only red circle in the set – despite the fact that neither circles nor red shapes are unusual. These two illustrations are instances of an *uncommon class combination* (ST-Va). The studies [181, 195] deal with this subtype.

A high-dimensional set may also constitute a so-called corpus, in which the individual cases represent different texts (e.g. documents, blog posts or e-mails). In this purely qualitative context the case’s word order is irrelevant for the anomaly’s description and detection. The anomaly may reside in unstructured or semi-structured documents, CSV files with a single message on each row, bag-of-words representations, or sets of a similar nature such as market basket data [100, 323]. A transaction consisting of an unusual combination of common retail products is an example of a market basket anomaly [196, 197]. Text cases such as blog posts or e-mails may be deviant because they contain unexpected topics or feature a different writing style [198-201]. See the ST-VIIIa subtype below for more details on text style and topic analysis. Dependent data afford wholly different subtypes. A tree, essentially an acyclic graph comprising qualitative identifiers of parent and child nodes, is a data structure well-suited for hosting Type V anomalies. One subtype in this context is the *deviant categorical vertex* (ST-Vb). An individual node in a tree can be anomalous as a result of its structural relationships. This requires at least the vertex ID (a qualitative designation identifying the individual nodes) and the edges (parent-child relationships). For example, a leaf node per definition is dependent on its structural relationships: there needs to be a parent, but children are absent. A leaf node that is deviant due to its graph context, e.g. because it is the terminal node of an extremely short path to the root, is therefore an instance of an ST-Vb anomaly (see Fig. 7.A). Other examples are vertices with an unusual amount of children (see Fig. 7.A as well) and vertices connected by an edge that has unexpected labels [cf. 132]. Note that a vertex with a single uncommon categorical value is simply a Type II anomaly, since no dependent data are required to describe and detect it in a flat node list. An ST-Vb anomaly is not necessarily a node in a tree. It can also be part of a regular graph, assuming that weights or other numerical properties are not involved in the anomalous behavior. An example is a vertex that is entirely unconnected (see "A.F" in Fig. 7.B) or does not belong to an identifiable community [202]. A node can also be deviant because it is connected to an extreme amount of other vertices, which in various domains is known as a 'super spreader'. In biology this refers to a single individual who disproportionately infects a large number of other people and as such contributes to the speed and degree of the outbreak, a phenomenon observed for Covid-19 and other viruses [154-156]. Likewise, in a security context this may be an infected source node in a computer network that communicates with many other nodes, possibly with malicious intent [203, 296]. Graphs are well-equipped to deal with notions of locality by taking into account adjacent nodes or the broader community. This allows anomalies such as a vertex with a class label that is unexpected at that position in the graph [20, 204, 295]. Examples are a smoker in a group of non-smokers and vertex "B.X" in Fig. 7.B (seemingly mislocated in the "A" group). The connectedness of neighboring nodes can also be analyzed [205]. An anomalous vertex then is a node whose neighbors are all highly connected or mostly unconnected. A node connected to two otherwise separated graph communities, such as vertex "A.E" connecting the "A" and "B" groups in Fig. 7.B, can also be seen as an anomaly [20, 206]. In real-world data such community-crossing occurrences may point to intrusion attempts [207]. A related subtype is the *deviant categorical edge* (ST-Vc). Many of the examples provided above for the ST-Vb subtype have an analogue for the deviant categorical edge. Examples are a relationship that connects two otherwise separated communities [cf. 20, 206], a hyperlink between two web pages with unrelated information [324], and a link that is attached to a vertex with an uncommon class label.

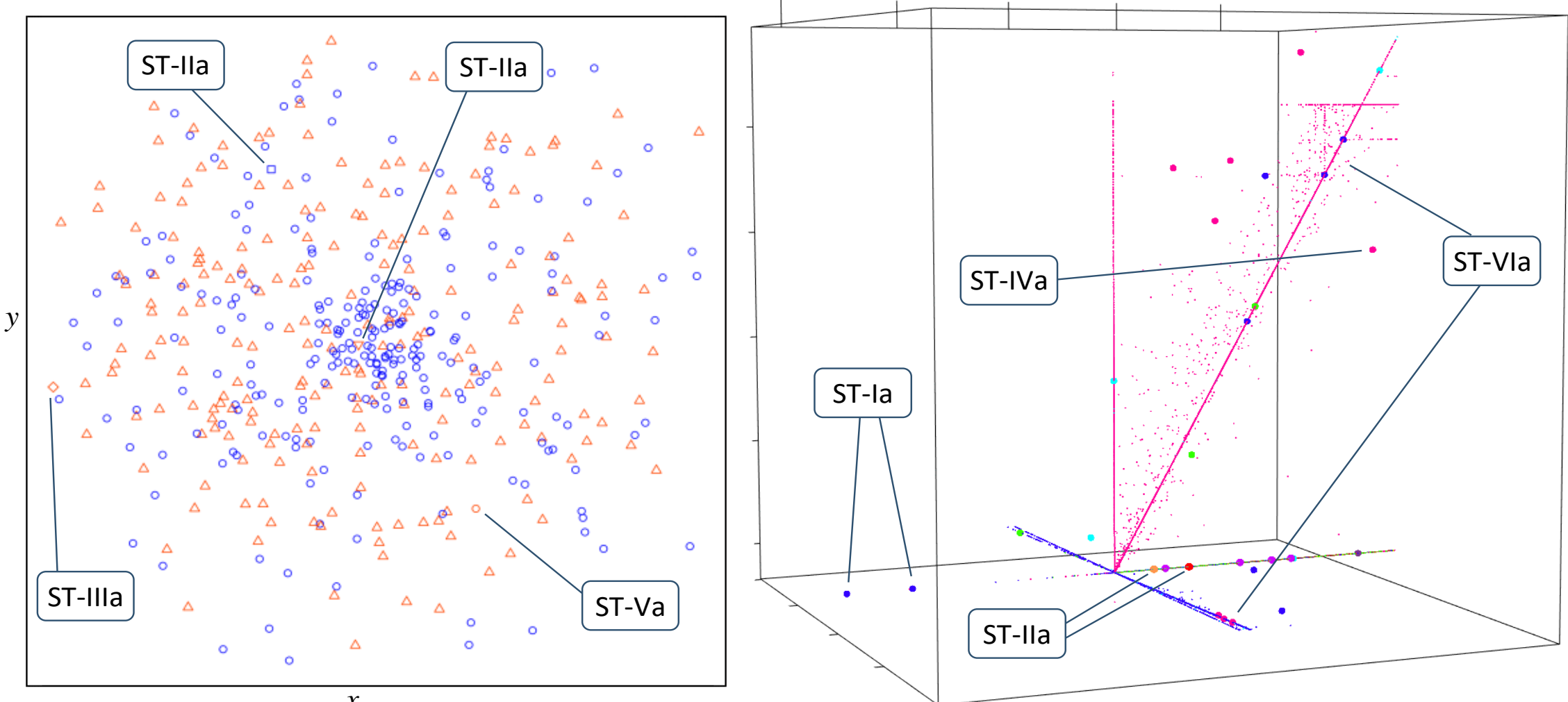

**Fig. 6: (Left) Synthetic set with two numerical attributes and two categorical attributes (color and shape); (Right) Real-world Polis set with one categorical and three numerical attributes, and large dots being anomalies.**

**VI. Multidimensional mixed data anomaly**: This is a case that does not fit the general patterns when the relationship between multiple quantitative and qualitative attributes is taken into account, without being an atomic univariate anomaly with regard to any of the individual attributes that partake in this relationship. It concerns a case with an unusual combination of qualitative and quantitative attributes, which can reside in both independent and dependent data. As with all multivariate anomalies, multiple attributes need to be jointly taken into account to describe and identify them. As a matter of fact, multiple data types need to be considered, as anomalies of this type per definition are comprised of both numerical and categorical variables.

In a set with independent data the anomalous case generally has a class value, or a combination of class values, that in itself is not rare in the dataset from a global perspective, but is only uncommon in its own neighborhood. Such cases therefore seem to be mislabeled or misplaced. The *incongruous common class* (ST-VIa) is such an anomaly and this subtype has been described in various studies [6, 70, 160, 208]. The right panel of Fig. 6 shows several real-world ST-VIa occurrences identified in a data quality analysis of the Polis administration, with multiple blue and pink dots seemingly misplaced or mislabeled. Not all detected anomalies necessarily represent erroneous data, as complex real-world phenomena sometimes simply result in strange (but correct) data. However, this specific analysis showed that some occurrences proved to be indicative of real data quality problems, which were subsequently remedied by improving the software [6]. Cases *ST-VIa* in Fig. 4 are also examples in this administration, as they are data points with a color (class label) rarely seen in their respective neighborhoods.

In dependent data a Type VI anomaly can manifest itself in many other ways. An *incongruous common sequential class* (ST-VIb) is an individual deviant in a sequence of class values of one or several substantive attributes. A quantitative time or sequence indicator is required here to link the dependent substantive values, although in symbolic data the order may be implicit. An example is the red underlined class `phaseB` at an unexpected position in this symbolic sequence:

phaseA, phaseB, phaseC, phaseA, phaseB, phaseC, phaseB, phaseA, phaseB, phaseC, phaseD, phaseA, phaseB, phaseC

(Note that the blue underlined case `phaseD` is a Type II anomaly because it is an entirely novel class.) Another example can be found in a DNA segment. This is a symbolic sequence in which each of the characters represents one of four nucleotide bases, namely A, G, T or C [10, 57, 183]. After reading the data the individual characters of the genome sequence are automatically verified and corrected in order to obtain a complete and accurate representation of the chromosomes. The order of the base symbols herein contains information that can be used for this verification and correction task [121, 209]. The characters in this example constitute qualitative data, but the substantive information in ST-VIb occurrences may also consist of mixed data. In fact, the verification process that determines the quality of the DNA character reading often also utilizes the underlying quantitative chromatogram data that are available for a given base [ibid.]. Another example is when a numerical sequence variable, substantive numerical variables (e.g. 'amount of money') and categorical variables (e.g. 'type of transaction') form a time series that hosts anomalies. The blue dot in Fig. 10 is an example in crop biomass data, which may indicate a wrong label of the data point.

A graph, comprised of numerical weights and nominal vertex IDs and edge directions, is also a structure capable of hosting Type VI anomalies [cf. 112, 113, 148, 149]. In this context an anomaly can take the form of a *deviant vertex* (ST-VIc). A specific example is a node connected by multiple edges with high weights, which may be of interest because such a vertex potentially has a high impact in the network. From a security perspective such an ST-VIc node could be infected if it sends many packages or may be subject to a DDoS (distributed denial-of-service) attack if it receives many packages from a great number of sources [152, 210]. From a technical perspective a node with many high weights may also point to faulty equipment [205]. A node with exactly one very heavy link constitutes an anomaly as well, which in a who-calls-whom network could indicate a stalker who keeps calling one of his or her contacts [ibid.]. (Note that the edge with the overly high weight, such as shown in Fig. 7.B, is itself an ST-Ia anomaly because it simply is an extreme value in a weight vector.) Attributed graphs readily afford the detection of local anomalies, i.e. vertices featuring substantive values that differ from their neighbors [211]. An interesting example can be found in a graph representing individual people (vertices with attributes such as monthly income) and their friendships (edges that connect people). A person with an income below average and connected only to rich people will likely be a rare occurrence [169]. A *deviant edge* (ST-VIg) is another subtype of the Type VI anomaly. An example is a link with a weight that can be considered normal in the entire graph but is relatively high or low in the local community or subgraph, such as the ST-VIg example in Fig. 7.B. In a communication network such edges may point to fake or redundant message exchanges between vertices [152].

In biology a phylogenetic tree represents the evolutionary relationships of species, individuals or genes from ancestors to descendants [106, 212, 213]. Such a tree shows that the brown bear and the polar bear evolved relatively recently from a common ancestor, while the split between these two species and the giant panda occurred closer to the root, thus in a more distant past. Although the topology can be represented with purely categorical data, biologists often use a mixed data tree in which the branch lengths (edge weights) represent the genetic distance. An individual tree branch (edge) can be anomalous if it is significantly shorter or longer than the other branches in that neighborhood of the tree, and as such may point to an interesting difference in the evolutionary rate of species or to a methodological problem [212, 214-219]. Individual branches may also be anomalous because they are unstable, which means these edges easily jump to very different positions in the tree. This may be observed when comparing trees generated with different samples or by different tree-building algorithms, or when the parameters are slightly changed [215-217].

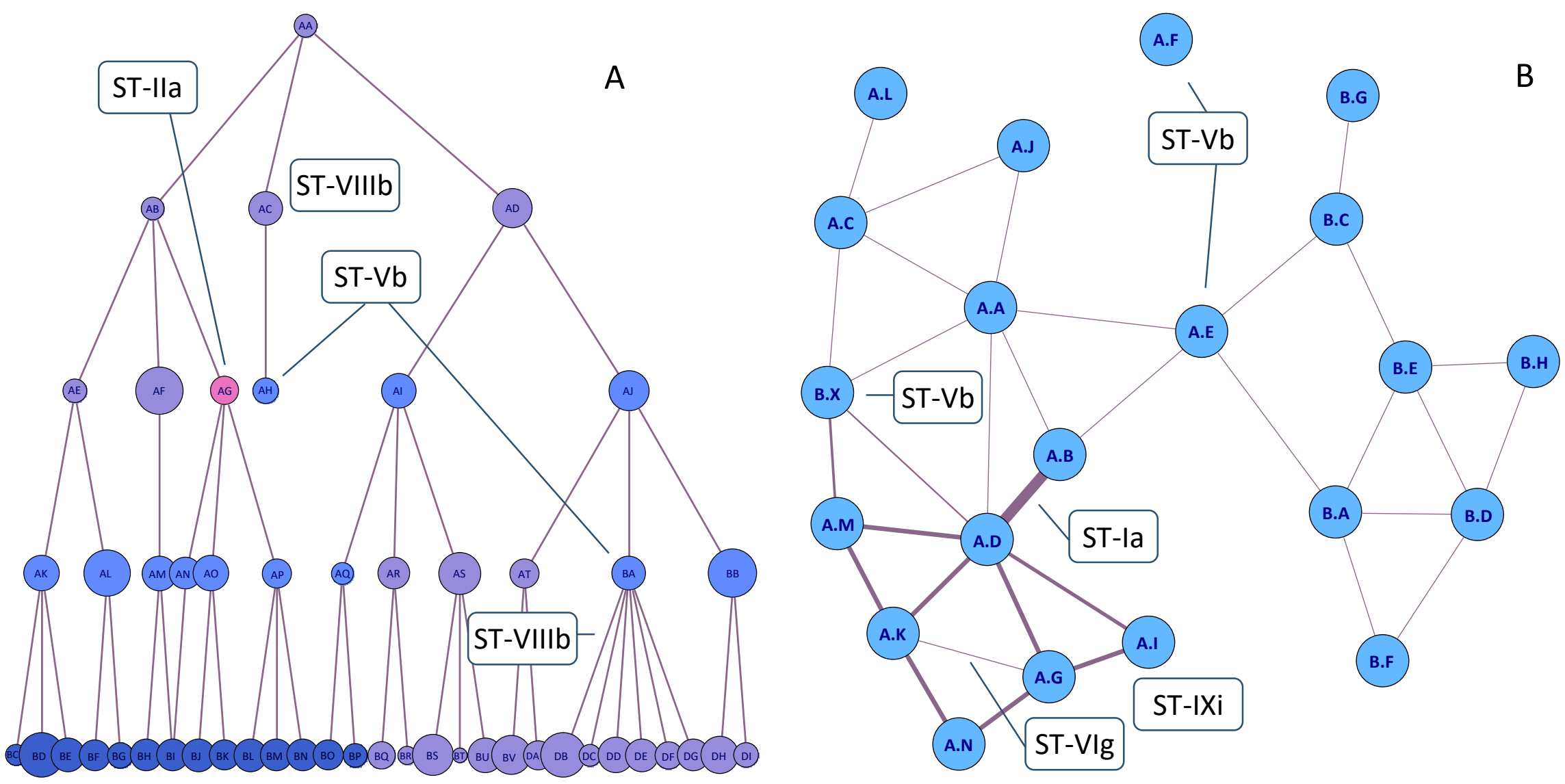

**Fig. 7: Various types of anomalies in (A) a tree and (B) a cyclic graph.**

When dealing with mixed data the focus of study can also be a dynamic graph, in which time-dependent behavior is taken into account [20, 112, 113, 149]. This may take the form of irregular changes in the respective time series, such as spikes or level shifts with regard to the edge weights, attribute values, or the frequency of vertices and edges [20, 112, cf. 204, 220]. See the different sequence-based subtypes for a full overview of how these dynamics can manifest themselves (i.e. ST-IVe and ST-VIb for atomic anomalies, and ST-VIIa-f and ST-IXa-i for aggregate occurrences). Notable events in this context can also be graph-specific, in the form of *unusual insertions, modifications or removals of vertices* (ST-VId-f) and edges (ST-VIh-j) in the network [112]. An example is an 'evolutionary community outlier', i.e. a node whose time-dependent behavior is different from that of its neighbors or community members [221]. A node that at some moment in time switches in terms of community membership can be regarded as an ST-VIe anomaly [112].

A *deviant spatial point* (ST-VIk) is a case with coordination data, often in combination with substantive properties, that can be seen as unusual. Although deviations in this context can be global [279], the explicit coordinates can be naturally exploited to define neighborhoods and detect local anomalies. For Type VI cases this typically concerns geographical sets, which are known to generally involve mixed data [126-128, 222]. An ST-VIk anomaly usually represents a unit location with one or more properties considered to be abnormal in that spatial neighborhood [211, 222, 223]. Examples that combine labels and the point's area are a dam in a residential area and a car in the middle of an ocean [cf. 316]. When a time dimension is also present the set may host *deviant spatio-temporal points* (ST-VIl). These are cases with one or more values that seem unusual when both their temporal patterns and neighboring points are taken into account [277, 283]. For example, a case's temperature, wind direction (e.g. "NNW") and wind speed – measured at a given time and location – can be unexpected in the context of the historical data of that geographical area [128, 224]. Spatio-temporal anomalies like this have been shown to point to complex climatic phenomena such as El Niño [128, 184, 225].

### 3.2.3 Aggregate anomalies

This section provides an overview of aggregate anomalies. Such an anomaly is a group of related cases that deviate as a collective. The cases are generally not individually anomalous, but multiple cases are jointly involved in a deviation from the dataset's regular inter-case patterns that can be expressed in terms of several qualitative and/or quantitative variables.

**VII. Aggregate numerical anomaly**: This is a group of related cases that deviate as a collective with regard to their quantitative attributes. Such an anomaly is typically found in time series data, in which it constitutes a subsequence of the entire sequence. A time series is capable of hosting a variety of aggregate anomaly subtypes, the discovery of which is strongly related to the task of change detection. A first subtype is the *deviant cycle* anomaly (ST-VIIa), illustrated in Fig. 8.C. This subtype occurs when the time series consists of cycles – such as climatic seasons – that demonstrate similar patterns, with the discord following a different pattern [7, 16, 69, 189]. Deviant cycles can be observed in many natural and societal phenomena, and have also been shown to correspond to unexpected physical gestures that were originally represented as video imagery [114]. The cycles can generally be detected in an unsupervised fashion, but for very specific deviations, such as certain medically relevant heartbeat patterns in an electrocardiogram, a supervised approach may be required [226].

Another subtype is the *temporary change* (ST-VIIb), which is a rise or fall of the substantive value that requires a certain period to get back to the regular level [138, 140]. This subtype has many practical uses, such as detecting wild fires and other ecosystem disturbances [304]. Fig. 8.B shows an example of an abrupt change that gradually returns to the normal range, a real-life example of which is the burst in the volume of news articles following an earthquake, major crime or other dramatic event [227, 228]. Another example is a so-called 'transient' in audio data, i.e. a sudden increase in amplitude followed by a decay [123, 229]. Examples of this are a gunshot and a microphone being dropped on the floor. Although the initial spike can be identified as one or several atomic extreme value anomalies, the slowly diminishing tail is also an intrinsic part of the anomaly. It renders this a collective that can only be described and detected as a combination of multiple cases and attributes. A traditional temporary change anomaly is often described as abruptly starting and gradually returning to the regular level [138, 140, 315]. However, the ST-VIIb subtype in this typology also allows gradual starts and sudden recoveries so as to accommodate more variations, akin to how the 'patch outlier' of [187] and the 'spike' and 'stuck-at' deviations of [314] can manifest themselves. An interesting astronomical example can be found in the light curves that represent the brightness of stars. Fig. 9.A shows the time series of the star WASP-47, based on publicly available observations from NASA's Kepler space telescope [250]. The sudden dip in brightness, which extends over several data points, can be explained by various phenomena. In this case it represents a planet transiting the star, and such dimming events thus offer a way to discover exoplanets [230, 231]. Note that this exhibits only a 1% dip, which requires these measurements to be very precise.

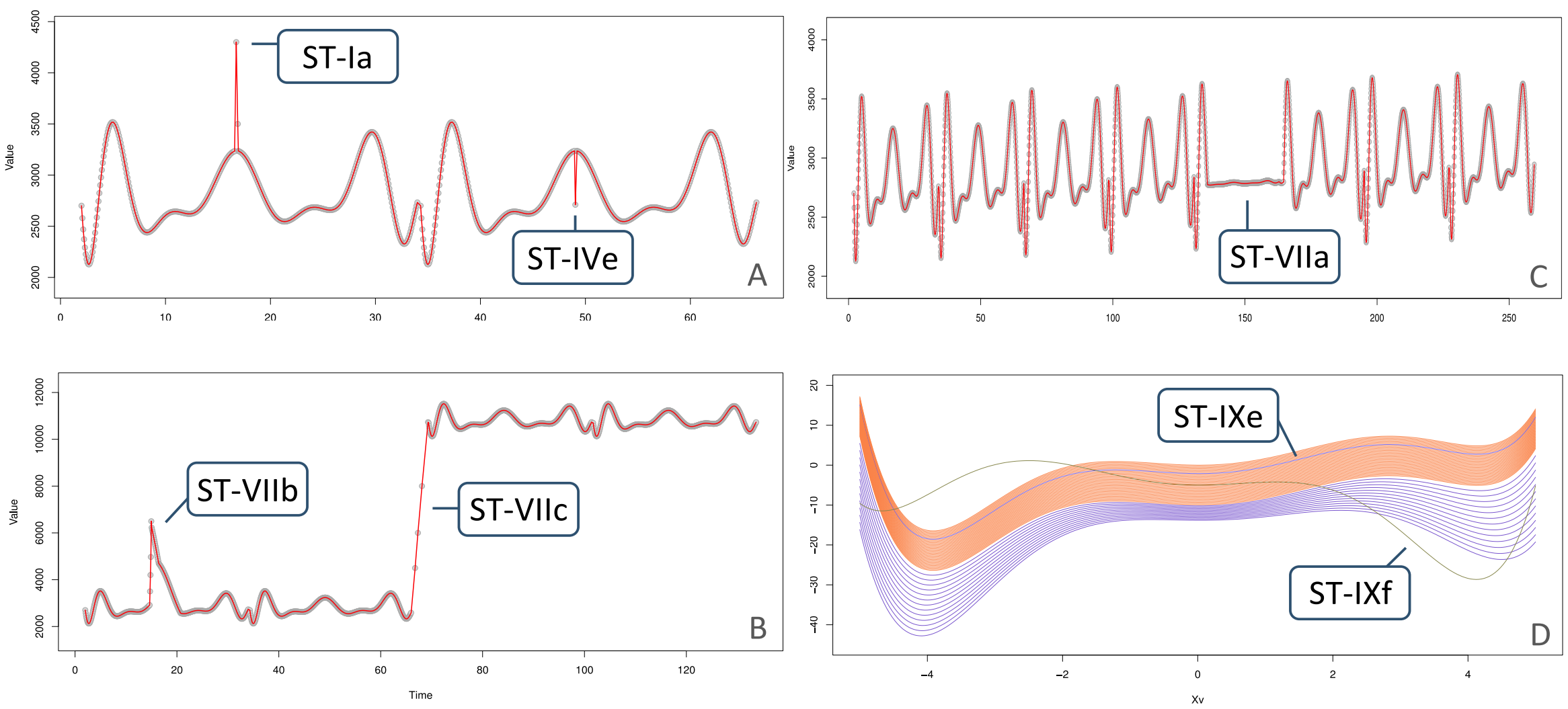

**Fig. 8: Time series and panel data anomalies. Gray dots represent individual measurements, red lines show temporal dependencies.**

Another subtype of a Type VII anomaly is the *level shift* (ST-VIIc), which is an abrupt structural change to a higher or lower value level, i.e. a permanent step change [138, 140]. Such a shift, illustrated in Fig. 8.B, comprises at least two consecutive data points and should therefore be regarded as a dyad or a larger group. A variation of this subtype, the 'seasonal level shift', implies a step change of the level of the recurring cycles or their amplitude [140, cf. 188]. An *innovational outlier* (ST-VIId) is a complex anomaly that goes beyond a temporary change, and usually consists of an initial shock followed by an impact on potentially both the cycle and trend components, and may be temporary or permanent in nature [21, 138-141]. A *trend change* (ST-VIIe) represents the start or end of a structural trend [232, cf. 184, 233]. Finally, the *variation change* (ST-VIIf) is an anomaly in which the random fluctuation changes over time [313, 329]. For example, the variance that was observed earlier may suddenly increase or drop to zero, possibly due to sensor failures, low battery levels or environmental circumstances that exceed the hardware range [314]. Note that in practice the variation change may be observed simultaneously with a temporary change or level shift, so in certain situations it may be worthwhile to acknowledge such a combined subtype separately and add it to the typology.

When the data exhibit nonstationarity or drift there may be an overlap with the aggregate time-dependent anomalies. Level shifts, trend changes and possibly innovational outliers are associated with a permanent alteration of the time series. These are related to the types of change in the underlying distribution that are also acknowledged in the literature on concept drift [234].

Spatial and spatio-temporal data comprise a set of coordinates, a set of numerical substantive features and possibly a time dimension, and often manifest themselves as images and videos. Although sometimes residing at the data point level, an image or video anomaly usually represents a region-of-interest, i.e. an aggregate rather than an individual pixel or voxel [67, 68, 122, 194, 235, 281]. A *deviant numerical spatial region* (ST-VIIg) is a quantitative aggregate that is unusual due to its spatial and possibly substantive features, e.g. a surprising object, area or person in a static image. Due to the spatial context local anomalies can readily be acknowledged. For example, the small blue region at the right of Fig. 9.C is anomalous with regard to its neighborhood but normal from a global perspective. Other occurrences may feature an unusual texture in an image, such as the different surface pattern to the south-east of the blue region. Identifying deviating textures has practical applications, including detecting scratches, dents and other sorts of damage during production processes in the manufacturing industry [312]. Nuanced high-level semantic anomalies, such as detecting a cat-like dog amongst multiple cats, can also be detected, although this may require prior theory, rules or supervised training because the needed domain knowledge cannot be leveraged otherwise [13, 310, 322]. A *deviant numerical spatio-temporal region* (ST-VIIh) is similar but also takes the time dimension into account [277]. This pertains to an area of the frame that is significantly different between images or video shots [194]. The identified differences between images or shots can reflect e.g. moved objects, altered light sources, transformed colors and changed camera positions. The region of the image change can also cover the whole frame, e.g. when it is significantly brighter or inserted as a subliminal shot. [68, 122, 235, 236]. Deviant spatio-temporal patterns also figure largely in the video surveillance of streets, parks and train stations. In non-crowded scenes anomalies may be an individual person or vehicle that demonstrates abnormal walking, running, crawling, driving or stopping behavior [68, 293]. In crowded scenes individual people cannot easily be distinguished, so the focus is on even more aggregated and abstract motion patterns that capture multiple subjects simultaneously [237]. Examples are pedestrians that move against the general flow, an empty local area that is normally crowded (e.g. a ticket gate during rush hour), and groups of people that obstruct traffic.

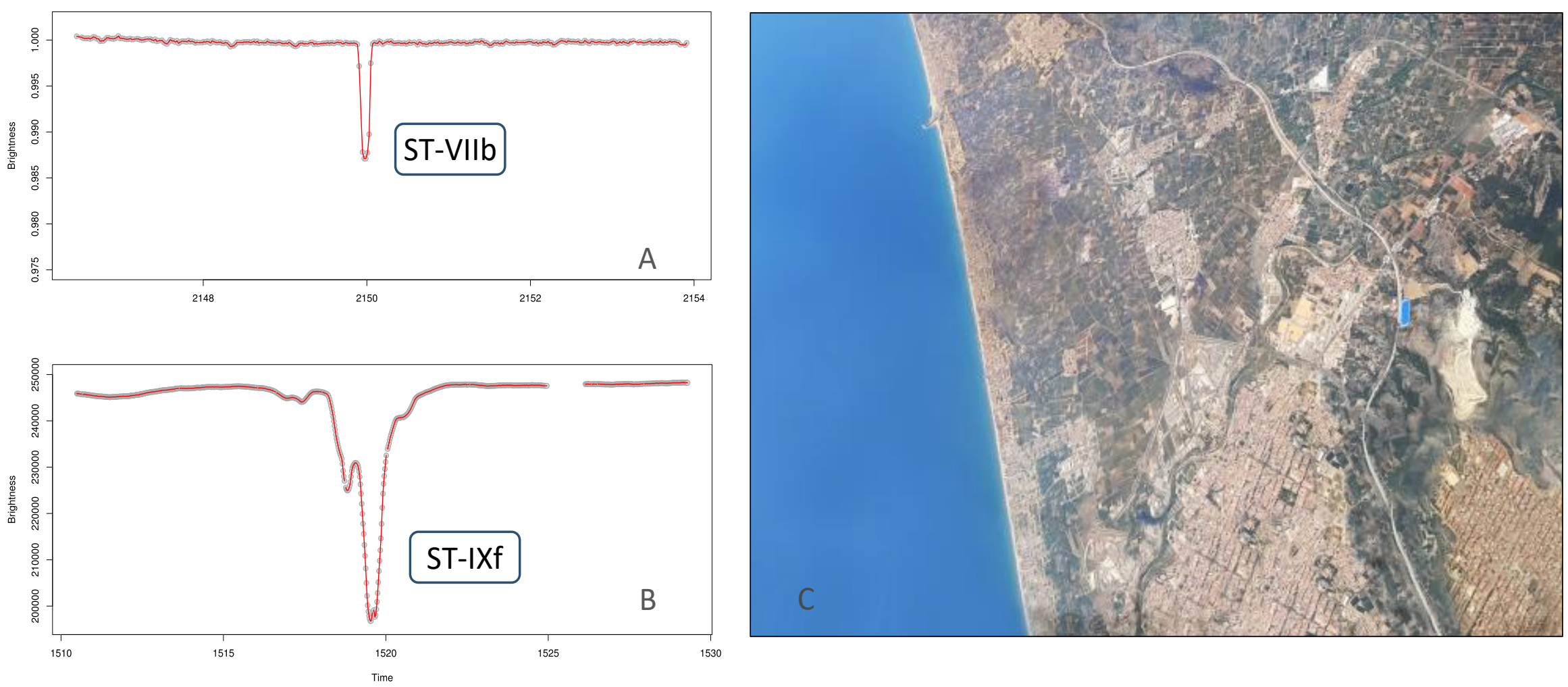

**Fig. 9: Real-world data: A and B are measurements from the Kepler space telescope. C is an aerial photo.**

Aggregate anomalies are mostly found in dependent data. However, research has shown that they may also occur in i.i.d. data [288-291]. The first subtype in this context is the *point-based aggregate anomaly* (ST-VIIi) and simply consists of a set of anomalous data points, i.e. a micro cluster of points that are individually deviant. The current typology positions them as an aggregate because the literature regularly discusses this subtype as such [e.g. 288, 290, 291]. However, in certain contexts it may be more appropriate to treat them as Type IV cases. The second subtype is the *distribution-based aggregate anomaly* (ST-VIIj), which necessarily is an aggregate anomaly because its deviant nature is dependent on group-level characteristics. A dataset may consist of different clusters, with the anomalous cluster exhibiting e.g. a deviant variance, covariance, mean or group frequency [282, 290, 291]. The scagnostics approach also acknowledges relevant distributional metrics regarding density and shape, such as to what degree a collection of data points is striated, stringy, skewed and clumpy [307, 308]. An example of a density-based VIIj anomaly is when it takes the form of an excess density pattern on top of the normally observed distribution [289].

**VIII. Aggregate categorical anomaly**: This is a group of related cases that deviate as a collective with regard to their qualitative attributes. A *deviant class aggregate* (ST-VIIIa) often manifests itself as a deviant text paragraph, section or document. An individual symbolic segment in this context is not conveniently stored in a neatly structured format (as with Type V occurrences, with one case being a single document, e-mail, transaction or blog post). The segment here exists as an aggregate concept, e.g. as multiple rows or sentences in an unstructured text or bag-of-words, or as multiple sections in a semi-structured text. For example, an aggregate of transactions in a market basket set, grouped over e.g. time, region or client, may be anomalous. An ST-VIIIa anomaly may also be a novel text that differs with regard to topics and content [100, 198, 227, 238]. Similarly, a text may stand out in terms of style or tone [199, 201, 239]. This has practical applications because a segment with a deviant style may point to plagiarism [200]. Examples of textual and distributional features relevant here are the percentage of words of a given type (e.g. pronouns, adjectives, nouns, prepositions), the ratio of adjectives to nouns, the average number of syllables per word, spelling mistakes, the ratio of positive versus negative words, and the percentage of words that occur only once (note that the unique word itself is a Type II anomaly). These characteristics, which do not require word order, can be used to identify the most dissimilar texts.

Similar to Type V cases a tree or other graph is a structure suitable for hosting Type VIII anomalies. A *deviant categorical subgraph* (ST-VIIIb) is a subset of a graph and is unusual in terms of its qualitative attributes. A typical manifestation is a subtree, which comprises multiple vertices and edges. Various characteristics of the dataset may underlie the anomalous structure, such as an unusually long or short path from leaf node to root, or an otherwise uncommon path in the collection of paths. Fig. 7.A features two examples, namely the unusually short path "AA-AC-AH" as well as vertex "BA" in combination with its unusually large number of descendants. An uncommon split in a tree should also be regarded as an aggregate anomaly, which comprises at least two edges and three vertices. In addition to these topological aspects, the anomaly may involve substantive attributes. A subgraph that deviates from a regular pattern of domain-specific class labels is an example of this, such as linked vertices of classes that normally are observed to be unconnected [18, 204].

A *deviant relational aggregate* (ST-VIIIc) is an anomalous occurrence of a non-atomic concept represented by its relationships. Such complex collectives are usually comprised of several domain-specific entities that are inter-related by one-to-many relationships, and are typically stored in different tables in a relational database [95, 124, 125]. The structural relations between the entities are modelled using nominal ID or key attributes, which can demonstrate patterns and deviations [19, 95, 240]. Similar to graphs, the topological structure alone – thus not taking into account any substantive attributes – allows for patterns and anomalies. For example, if normal cases relate to one or two relationship-instances of another entity, an occurrence with a large collection of such instances can be considered an anomaly, such as a person with seven jobs in income data. Also, suboptimal data models and systems may lead to several well-known database issues, namely insertion, deletion and update anomalies [125]. An example is an 'orphaned record' whose foreign key relation points to a previously deleted record in the primary table [241]. Substantive categorical attributes may also play a role. For example when a case from a given entity type (stored in database table A) is unlike regular occurrences because it has an unexpected class value in a related entity type (stored in table B). A concrete example is the sales transaction of an item that at that moment had the unusual status of being out of stock. What is 'expected' can be based solely on patterns in the given dataset, but in practice often also on the data model, domain-specific rules or a supervised approach. Note that if numerical attributes are also part of the definition of the aggregate deviant, it will be a Type IX anomaly (to avoid redundancy this variant is not explicitly described in the next section).

**IX. Aggregate mixed data anomaly**: This is a group of related cases that deviate as a collective with regard to both numerical and categorical variables. Owing to its mixed data and diverse ways in which relationships can manifest themselves, this type allows for a wide variety of complex anomaly subtypes. To start, an anomaly can be a deviant group of class values within the larger symbolic sequence of one or more attributes [5, 7, 85, 114, 183, 242]. Such subsequences can for example reside in DNA strings, system logs or time-based text messages. Order or time information is required to link the consecutive data points. A first subtype in this context is the *class change* (ST-IXa), in which a sequence suddenly changes its symbolic value or pattern. This is akin to the aforementioned level shift in purely numerical data (ST-VIIc) and comprises at least two consecutive cases. A simple example is when the function of a given real estate asset is at some point changed from 'COMMERCIAL' to 'RESIDENTIAL'. A more complex manifestation occurs when the stochastic univariate or multivariate pattern of the given categorical attributes changes. This could be observed in a time series of market basket data when two or more products are increasingly bought together, with this pattern being non-existent in the past [192]. A third example of the ST-IXa subtype is akin to the temporary change, and can be observed when one or multiple news streams suddenly introduce a new topic, perhaps as a result of a major disaster or crime [227, 228]. (Note that the example of bursty news topics presented above as a Type VII anomaly simply pertains to the time-related *frequency* of news items on a given topic, whereas the current example focuses on the *topics* themselves, i.e. as a pattern of words.) In the real estate and market basket examples the underlying distribution of the dataset changes, meaning that it is subject to some form of data drift [190, 192, 234].
Another subtype that may be found in sequence data is the *deviant class cycle* (ST-IXb), in which an entire subsequence is anomalous because it does not adhere to the cycle pattern. The pair of red underlined classes in the following phase-sequence can therefore be regarded as such a subtype:

phaseA, phaseB, phaseC, phaseA, phaseB, phaseC, phaseA, phaseC, phaseA, phaseB, phaseC, phaseB, phaseA, phaseB, phaseC

(Note that the blue underlined phaseB case is an ST-VIb anomaly because it is a known single class at an unexpected position.) Such anomalous n-grams or other forms of subsequences are typically of interest in (intrusion) detection systems, where deviations from regular symbolic sequences may indicate a defect or attack [60, 85]. A concrete example is when the short login-password event, which usually is observed once every now and then, suddenly occurs very frequently in a row and as such may represent a hacking attempt [183, 275]. An example from a different domain is a DNA sequence that contains genes from another organism [243]. A more intricate variation of the ST-IXb subtype combines, apart from the sequence attribute, substantive numerical variables with categorical variables [244]. An interesting example can be found in crop biomass monitoring, in which vegetation quantities (represented by the NDVI index) and crop classes are included in a single time series [292]. Cycle 14 in Fig. 10 is an example, with the anomaly possibly pointing out a data quality problem because the cycle's crop labels may very well be erroneous. A symbolic sequence can also be anomalous in its entirety when compared to other sequences. A real-life example of such a *deviant class sequence* (ST-IXc) would be an anomalous trace in a collection of traces representing logged system or business processes [85, 183, 294]. Another example can be found in a dedicated genomic database that contains an erroneously included DNA sequence from an entirely different organism [243]. Finally, subtype ST-IXc may represent an anomalous text paragraph, section or document. This is similar to the topic and style deviations of ST-VIIIa, although the order of the words now plays a crucial role. For example, texts may deviate in terms of sentiment, which is a relevant issue in detecting fake reviews [157]. Word order is especially important here to properly deal with negations and other types of sequential dependencies [157, 245, 246, 299].

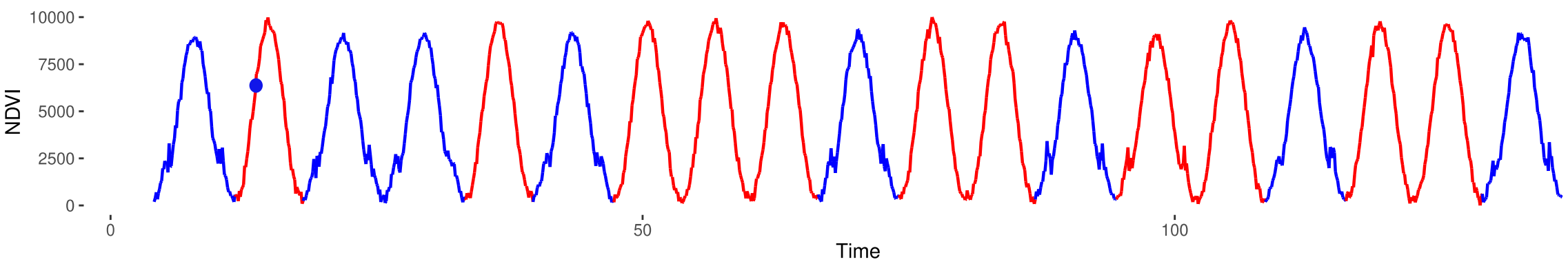

**Fig. 10: Crop biomass time series, with color representing the class of crop. The large dot in cycle 2 highlights an atomic anomaly, i.e. a data point with an unexpected class label, whereas cycle 14 is an aggregate anomaly.**

Several related subtypes are unusual individual time series in a panel dataset hosting multiple time series [116, 247, 282]. The set's substantive information consists of numerical (and possibly categorical) data, whereas the individual series are distinguished by a nominal attribute [57, 111]. The *deviant sequence* (ST-IXd-i) consists of six subtypes [142, 233, 280, 314]. The anomalous series may be *isolated* from the other series for a short time interval or an extended period. In case of the latter, the curve may be deviant due to a *shift* (having a normal shape but at a different location), a distinct *shape* (positioned at the same location as other curves but showing another shape), a different *amplitude* (having the same shape but with a different range), an unusual *trend* (that makes the sequence slowly drift away from the other sequences) or with different *variation* levels (exhibiting a different random fluctuation). Fig. 8.D shows an ST-IXe anomaly that has a deviant shift with respect to its categorical property represented by a blue color, as well as an ST-IXf occurrence with a deviant shape. Another example is a deviant sound recording in a music collection, such as a song that stands out in terms of loudness, rhythm, melody, harmony or timbre [248].

The field of astronomy also offers an intriguing example with the star KIC 8462852. When analyzing its light curves as measured by the Kepler space telescope one can observe brightness dips, a type of event that was declared earlier in this study a temporary change anomaly. However, to understand exactly why astronomers where so excited about KIC 8462852 it is necessary to view the anomaly in the context of the usually observed dips. Fig. 9.A represents a normal occurrence, i.e. a fluent dip of around 1% that lasts a couple of hours, which may very well point to a transiting exoplanet (which is indeed the case here). KIC 8462852, on the other hand, exhibits several irregular, aperiodic brightness fluctuations that last multiple days, with strong light dips of up to 22% [249]. Fig. 9.B shows this anomaly, based on publicly available real data from the Kepler telescope [250]. Compared with the normal dips of other stars this is thus an ST-IXf *deviant shape* sequence due to its erratic and extended dimming. The cause of this anomalous event is unknown, but the fluctuations cannot be attributed to the light being blocked by a planet orbiting the star. Hypotheses that have been put forward are an uneven ring of dust, a swarm of asteroids or comet fragments, and even the spectacular explanation of an artificial megastructure built by an advanced extra-terrestrial civilization, i.e. a so-called Dyson sphere for harvesting the star's energy [249, 251, 252].

Note that deviations with regard to seasonality, which besides the trend is another basic component of time series [111, 232, 253], are captured by the deviant shape and amplitude subtypes. Furthermore, the five ST-IXd-h subtypes can also be used to typify trajectory anomalies because time series are conceptually very similar to sets with information on moving entities [5, 116, 119, 120, 278, 325]. Examples are a ship that deviates from the normal route [254] and a taxi that takes an unusual amount of turns [316]. Beyond this, time series are able to represent visual shapes of physical objects. In this context sequences with a deviating shape have been used to detect skulls from a primate species that differs from the rest of the collection [114].

A graph dataset can host a *deviant subgraph* (ST-IXj), which is an anomaly comprised of multiple vertices and edges. A specific example is a group of linked vertices with significantly different substantive values than those observed in other parts of the graph. The anomaly can also reside in a set of domain-specific subgraphs, which are defined by a group ID or shared property. The deviation of the subgraph may be due to unusual structural relationships, weights or attribute values. Alternatively, a set of subgraphs may be derived by community detection, after which uncommon (sub)communities can be identified [95, 169, 255]. The subcommunity below left in Fig. 7.B with its high weights forms an example of such a phenomenon. The evolutionary trees used in biology also provide informative examples of graph anomalies. A so-called molecular clock may be assumed in such a tree, in which the branch length (edge weight) represents a time estimate for the evolution of one species into another [106, 213]. The evolutionary paths from the leaf nodes to the root need to have the same length because the evolution of a group is generally expected to have a constant rate. A given path from root to leaf that is significantly shorter or longer than the other paths can therefore be seen as an example of an anomaly [106, 212, 214, 218, 219, 256]. A clade (i.e. a subtree of an ancestor and its descendants) may also be considered anomalous if it is unstable when certain parameters are changed, features conspicuous differences in its branch lengths, or has several long branches – all of which may indicate methodological errors in the tree building process [212, 215, 216, 218, 257, 258].

In dynamic graphs a time dimension affords analyzing the evolution of graphs and subgraphs [20, 112, 220, 259]. Analogous to Type VI graph dynamics this may take the form of the aforementioned sequence-based anomalies such as spikes, level shifts and other changes (see the sequence-dependent subtypes ST-IVe and ST-VIb for atomic anomalies, and ST-VIIa-f and ST-IXa-i for aggregate occurrences). Such time-aware analysis of aggregate anomalies has been shown to be relevant for detecting faults in a network of application services, where monitoring nodes and edges at the individual level would not properly account for clusters and noisy traffic fluctuations [260]. Anomalous events can also be graph-specific, manifesting themselves as (sub)graphs that *appear, disappear, flicker, merge, split, grow, shrink* or demonstrate *eccentric* behavior (ST-IXk-r) [20, 112, 221, 261, 262, 287]. An example of a subgraph exhibiting highly eccentric behavior, which incidentally requires a rule-based detection approach, is a fraudulent group in a financial network that features specific trading ring patterns [263]. The group members first follow a 'blackhole' pattern by exclusively trading amongst themselves, and subsequently a 'volcano' pattern by selling the stocks – which by then have increased in price – to non-involved traders.
A *deviant spatial region* (ST-IXs) is an aggregate that is unusual due to its quantitative spatial and possibly substantive features. Such an anomaly is typically hosted by a geographical dataset with mixed data. An example is a deviant area, i.e. a polygon comprised of multiple lines, such as a land parcel with an unusual structure, color or class (e.g. water, development area or coastal scrub). Another example is a (poly)line located in an area where that class of object is normally not found, such as a river in the middle of a lake. These are examples that require not only linking different data elements to form polylines or other aggregates, but also relating them to both the wider area or polygon they are located in as well as any relevant domain-specific attributes. This is illustrated in [222, cf. 223] by counties (represented as polygons) in the United States, with Los Angeles and downtown Chicago being anomalies with an unusually high population density compared to adjacent counties. Spatial data can also be analyzed from a time perspective. A *deviant spatio-temporal region* (ST-IXt) is a polygon or other aggregated object with one or more substantive values that deviate when both its temporal pattern and spatial area are taken into account [279, 283]. Alternatively, it can be viewed as a temporal occurrence that shows a pattern unlike its spatial neighbors. In other words, it is a deviant (sub)sequence, such as one of the ST-IXc-i subtypes described above, but typically one that is uncommon in the local region rather than in the entire global space [224, 277]. An example is a region and time interval where the risk of suffering from a global disease outbreak is increasing significantly faster than elsewhere [281]. A more specific manifestation is a 'flow anomaly', a spatially marked subsequence that does not adhere to the pattern of values flowing from one location to another with a given time lag [283, 284]. Such occurrences may be observed if multiple sensors are placed in rivers and may point to flood conditions or chemical spills.
Finally, i.i.d. data with mixed data types may also host aggregate anomalies. A *point-based mixed data aggregate anomaly* (ST-IXu) is a group of cases that each have one or more unusual categorical and numerical values [290]. A *distribution-based mixed data aggregate anomaly* (ST-IXv) is a group of cases that is unusual in terms of its categorically and numerically determined group-level characteristics. An example is a cluster of neighboring points that is anomalous with respect to its distribution of categorical values [288]. See subtypes ST-VIIi and ST-VIIj for more information on their purely numerical counterparts.

To summarize, section 3 has introduced nine basic anomaly types, each of which is discussed in a tangible way by a variety of subtypes and ample real-world and synthetic examples. The basic types are stable due to the fundamental classificatory principles of the typology, while the set of subtypes is extensible.

## 4 Discussion

This section discusses several relevant topics, such as deciding on the anomaly type, the evaluation of AD algorithms, explainable anomaly detection and local versus global anomalies.

**Deciding on the anomaly type.** In order to sharply determine the nature of a given data deviation, one should opt for the simplest applicable (sub)type. For example, a case that is a Type II anomaly, per definition, will also have unique class value combinations in a larger subspace that includes additional categorical attributes. However, this does not imply one should see this instance as a Type V anomaly. After all, the deviation can be defined more parsimoniously. To be sure, this does not exclude the possibility that the case in question is also a Type V anomaly, but that should then pertain to a different subset of attributes.

There are several general principles to determine which (sub)types are simpler than others. The univariate types are simpler than the multivariate types. Anomalies that require only qualitative or quantitative attributes are simpler than those defined in terms of mixed data. Subtypes based on independent data are simpler than those based on dependent data, and atomic anomalies are simpler than aggregate types.
Simple subtypes such as the extreme tail value and unusual class may be part of a larger set with dependent data, e.g. one that constitutes a graph. The fact that the graph structure need not be referenced in the definition of these anomalies obviously does not prohibit the researcher to meaningfully discuss the deviation in the context of the given graph. Often it makes perfect sense to discuss a Type II occurrence as an "anomalous node", as long as one acknowledges that the graph structure itself is not required to define and detect the anomaly from a data perspective. The same holds for other complex data structures, such as spatial data and time series. This insight is the reason that the typology does not feature an anomaly subtype for a globally extreme high or low time series value. This is simply a Type I anomaly and, unless the position in the sequence should be identified, the time-related dependency with other cases in the dataset is irrelevant for defining and detecting the deviation.
This section concludes with a brief tightening of the terminology used in this article. The term *anomaly* and *deviant* are synonyms and used as general terms. The term *outlier* refers to numerically isolated occurrences, such as the classical Type I and Type IVa-d cases. The term *novelty* can also be defined more strictly, as referring to occurrences that in some way represent new and hitherto unknown events or objects.

**Evaluation of algorithms.** In addition to offering an understanding of the nature and types of anomalies, the typology is aimed at facilitating the evaluation of AD algorithms. This is a relevant contribution because most research publications do not make it very clear which types can be identified by the anomaly detection algorithms studied, nor do they position the targeted anomalies in a broader context. However, given the wide variety of anomalies, it is clear that individual algorithms will be incapable of identifying all types [6, 17, 60, 82, 84-86, 184]. Researchers can thus employ the predefined typology to provide a clear and objective insight into the functional capabilities of their AD algorithms by explicitly *stating which anomaly (sub)types can be detected*. Using the typology in this way also gives due acknowledgment of the no free lunch theorem [80-83] and demand for transparent and explainable analytics and AI [71-75] (also see next section).
However, the typology offers more opportunities for algorithm evaluation than merely clarifying which types and subtypes can be detected, since the typology is ideally also used for *creating test sets*. AD studies often evaluate algorithms by using existing AD benchmark datasets that are flawed [321]. Moreover, the common practice of treating (a sample of) a minority class as anomalies [9, 17, 86, 133, 159, 163, 182, 186, 195, 316] also poses problems. The cases of such a class may be very similar because they are part of a true pattern, and may even be very similar to other normal classes in the dataset. This latter situation can indeed be observed for several classes in the above-mentioned real-world Polis administration. In addition, there is no guarantee that all relevant anomaly subtypes will be present in such a test set. A better approach for creating AD test sets is therefore to use the typology as a basis for systematically creating and inserting instances of each relevant anomaly subtype in a real-world or simulated dataset. Such an injection approach ensures that the different anomalies are present and a thorough evaluation of the algorithm can subsequently be conducted. Researchers should aim to include the subtypes that, given the domain and typology dimensions, are relevant for the problem being studied.
Table 2 illustrates how the typology can facilitate the evaluation of several unsupervised algorithms for detecting anomalies in independent data. The focus is mainly on the algorithms, which are presented in the rows of the table, with the columns representing evaluation characteristics such as the capability to detect the individual anomaly types, metrics such as the AUC (area under the curve), F1 and sensitivity, as well as any remarks that may be relevant. Such an evaluation is appropriate when various (versions of) algorithms are evaluated and evaluation characteristics need to be presented at the level of the algorithm.
Grubbs' parametric test aims at verifying whether a vector contains one or two outliers [30, 264], while the related GESD procedure is intended for testing whether a vector contains multiple outliers [35]. SECODA is a non-parametric AD algorithm for mixed data [6, 70]. Distance-based algorithms use a nearest neighbor approach to detect anomalies [51, 52, 54]. One could also include several versions of the algorithms or pre-processing steps. For example, with distance-based AD techniques the numerical attributes could be normalized and the categorical attributes transformed using dummy variables or IDF (inverse document frequency), and such techniques could be evaluated in several combinations. Also, evaluation metrics based on the ROC (receiver operating characteristic) and confusion matrices could be calculated and reported for each anomaly (sub)type separately.
Table 3 presents the anomaly types in the rows, with the columns providing more details on different evaluation characteristics. This format was used in [70] for studying the impact of discretization and is appropriate if the focus is mainly on the anomaly (sub)types and few algorithms are being compared.

| Type<br>Algorithm | I | II | III | IV | V | VI | Remarks |
|---|---|---|---|---|---|---|---|
| Grubbs/GESD test | a: ✓ b: × | a: × b: × | a: ✓ b: × | a: × b: ×<br>c: × d: × | a: × | a: × | Also provides statistical significance metric. ST-IIIa will be detected using quantitative data only, and thus cannot be directly distinguished from ST-Ia. |
| SECODA | a: ✓ b: ✓ | a: ✓ b: ✓ | a: ✓ b: ✓ | a: ✓ b: ✓<br>c: × d: ✓ | a: ✓ | a: ✓ | No data type transformations or rescaling required, but vulnerable to the curse of dimensionality. ST-IIb cases are represented by high anomaly scores instead of low scores. |
| Distance-based AD | a: ✓ b: ✓ | a: × b: × | a: ✓ b: ✓ | a: ✓ b: ✓<br>c: × d: ✓ | a: × | a: × | Needs rescaling for optimal performance that corresponds with human intuition. With pre-processing (e.g. dummy variables or IDF) categorical data can also be analyzed. Type III occurrences cannot be directly distinguished from Type I outliers. |

**Table 2. Illustration of using the typology to evaluate anomaly detection algorithms.**

| Type | Impact? | Useful? | Explanation [ED = equidepth / EW = equiwidth discretization] |
|---|---|---|---|
| I | Y | N | ED cannot discriminate between univariate numerical values and is intrinsically not equipped to detect this type. |
| II | N/Y | Y | ED is identical to EW when analyzing a single categorical attribute. It can be more useful than EW if the goal is to detect (non-unique) rare Type II anomalies in numerically high-density regions in an analysis of mixed data. |
| … | … | … | … |
| VI | Y | Y | ED tends to favor the detection of Type VI anomalies and can be more useful than EW if identifying them is the aim of the analysis. |

**Table 3. Illustration of using the typology to evaluate anomaly detection algorithms, with the focus on the types.**

**Explainable anomaly detection.** The typology provides a generic framework for understanding anomalies through a data lens and is relevant for both research and practice. Typifying an individual deviant case as a specific anomaly subtype means it is described meaningfully in terms of five fundamental dimensions. This provides transparency and explains the nature of the deviation, i.e. makes clear how it differs from the normal cases in terms of key data characteristics. This adds value to an anomaly analysis because AD algorithms yield an anomaly score or label, but, although they may be able to detect multiple subtypes, typically do not provide information on how the individual anomalies differ from the rest of the data. The typology provides a structured way to analyze this and clarify the detection results.
For example, a general-purpose AD analysis on independent data, using e.g. distance or density based algorithms, may detect 12 subtypes, namely ST-Ia-b, IIa-b, IIIa-b, IVa-d, Va and VIa. These obviously represent a wide variety of deviations that may be present in the dataset. If the detection algorithm merely informs the analyst on whether (or to what degree) a case is anomalous, the results therefore remain a black box. The typology can be used to bring about the necessary clarification, with its 12 clearly distinguishable subtypes and names, as well as five dimensions that explain how each subtype differs from normal data. Fig. 4 illustrates different anomalies in a real-life dataset. Some of these occurrences differ from regular data because they have an extreme numerical value (ST-Ia) or have numerical values that, although individually normal, in combination position them outside the regular space (ST-IVa). Some cases are deviant because they feature a rare code (ST-IIa), while others have a code that, although normal, is unusual in their own numerical neighborhood (ST-VIa). By employing the typology these different subtypes can be clearly distinguished, labeled and understood. The plot also makes clear that visualization can be valuable during this explanation process and can be of help in typifying the detected anomalies.
Understanding the nature of the anomaly types present in a set is also relevant because of their different implications. In the context of data management AD may be conducted as an exploratory data quality analysis, which can be used to decide what kind of automated checks and corrections should be implemented. In real-time monitoring processes ST-Ia cases may then be handled with a simple threshold, ST-Ib cases with an interval, ST-IVa cases with a rule that combines thresholds or intervals, and perhaps it is decided that ST-IVc cases should be ignored because they merely represent subtle deviations.
The typology also yields tangible clues for further interpretation and sense-making. For example, a limited number of dispersed ST-Ia or b anomalies are likely to be non-informative random events or glitches, while the aggregate ST-VIIa, b, i and j occurrences may very well imply a more significant event or mechanism, and ST-VIIc and e even come with a drift in the data that represents a fundamental distributional change.

**Local anomalies.** There are several perspectives on the concept of *locality*, all of which can be meaningfully described in terms of the dimensions of anomalies that have been put forward in this study. The first perspective focuses on *data structures with independent data* and uses the *cardinality of relationship* to define the difference between local and global. Univariate anomalies are simply seen as global because they are unconditionally deviant relative to the remainder of the rows in the dataset [69, 70]. A single variable describes the entire (univariate) data space, which renders an unusual case in this view per definition a global anomaly. Therefore, when taking all the set's cases into account, Type I anomalies will always have an extremely low, high or otherwise unusual numerical value for the given attribute, without any condition and regardless of the other attributes. A similar argument can be given for Type II and III anomalies. The multivariate anomaly types, on the other hand, are only deviant given the categorical condition or the specific numerical area the case in question is located in. This is due to the fact that a multivariate anomaly is only unusual as a combination of values from multiple attributes and is therefore normal across the entire one-dimensional space. This is clearly illustrated with the mixed data types of the *ST-VIa* cases in the right plot of Fig. 6. These anomalies are normal with regard to the values of each individual numerical and categorical attribute. However, while e.g. the blue points are normal in the global space, they are deviant in their own local numerical neighborhood. A local anomaly thus exists in some area, subsegment or class of the data [cf. 278]. Other examples are provided by [57] in a discussion on global versus local anomalies. A male with a body length of 175 cm is normal in the general population, but is an anomaly within the local class of professional basketball players. The reverse is also true: someone with a body length of 195 cm may be unusually tall with respect to the population, but not when only considering the class of professional basketball players.
A second perspective on the concept of locality focuses on *data structures with independent data* as well, but uses the *data distribution* to make the distinction between local and global. Local anomalies are described in terms of neighborhood density or similar characteristics, a perspective that is particularly relevant if the set contains multiple clusters that differ in this regard [8, 17, 53, 55, 273]. The outlyingness of a single case can be seen as being dependent upon the degree of isolation relative to its local neighborhood (rather than to the global space). Techniques for this setting, such as LOF and LOCI, therefore determine the density of a case in relation to the density of its neighboring points.
A third perspective on locality focuses on *data structures with dependent data*. In such datasets the individual cases are by their very nature intrinsically related, allowing the data management or contextual attributes used for linking the data points to naturally define neighborhoods [8, 222, 223, 126, 128]. Local anomalies in this context are typically deviations from autocorrelation patterns. They can perhaps most clearly be illustrated with spatial data, in which latitude-longitude attributes or other sorts of coordinates explicitly define locations and neighborhoods. An occurrence is locally anomalous if the values of one or more substantive attributes are unusual in its own spatial neighborhood, but normal in other regions. This logic also pertains to images and videos, with data points being pixels or voxels with a fixed position in the canvas or frame. An example of a local anomaly in an aerial photo is an isolated tree in a certain region of the picture, while a large forest is present in another region [67]. The small blue region in Fig. 9.C is an example as well. A similar reasoning holds for time series data, in which the timestamp explicitly positions cases at time points and in periods, which are the temporal equivalents of spatial locations and neighborhoods. A local anomaly then has a value that is normal when taking into account the entire history, but unusual in the period in which the occurrence lies. The ST-IVe subtype is such an anomaly and is illustrated in Fig. 8.A. Local spatio-temporal anomalies can be viewed in the same vein. Finally, graphs also feature explicit structural positions and neighborhoods, although their visualization generally allows more freedom in where to graphically depict the nodes or cases.

**Other classifications.** The typology presented here offers an all-encompassing framework to describe the types of anomalies acknowledged in the literature, on the condition that they can be defined in terms of their data properties. For example, the typology can accommodate the well-known time series anomalies, i.e. the additive, temporary change, level shift and innovational outlier [138, 140, 141]. In fact, the typology makes a more detailed distinction, because the classic additive outlier type does not distinguish between outliers that are globally extreme and outliers that are only deviant from a local perspective, e.g. in their climatic season [141]. This study's typology identifies them as *extreme tail values* (ST-Ia) and *local additive anomalies* (ST-IVe) respectively. Another notable classification is that of [184], as it is both broad and data-oriented. However, it distinguishes between 9 concrete anomaly classes instead of 63 and does not use any classificatory principles.

The anomalies acknowledged in classifications that are not data-centric are not explicitly present in the current typology. For example, [96] presents classes of outliers that are not defined in terms of observed data characteristics, but instead refer to external causal phenomena that are often beyond the knowledge of the data analyst. This particularly holds for the procedural error (e.g. a data entry mistake), the extraordinary event (e.g. a hurricane) and the extraordinary observation (a non-explained measurement). Other classifications seemingly refer to the data, but are ultimately grounded in phenomena external to the dataset. Causes of outliers presented in [136] are, e.g., measurement errors and data from other distributions. To ascertain whether this is the case, one often requires additional information and subjective interpretation [2, 4, 34, 323]. This is by no means to say that these conceptualizations are not valuable, because analysts should certainly possess knowledge of the potential causes of anomalies. However, this study defines anomalies in terms of observable data characteristics and five theoretical dimensions. It allows the objective and principled declaration of anomaly (sub)types in a tangible and explainable fashion, using the data at hand and leaving relatively little room for discussion and doubt.

The classification in [7] distinguishes between point, contextual and collective anomalies. This differs from the three broad groups of this study's typology (i.e. atomic univariate anomalies, atomic multivariate anomalies and aggregate anomalies), which are rooted in the requirements a typology brings with it. The former classification features classes of anomalies that are not mutually exclusive, because a collective anomaly can also be a contextual anomaly. This is an undesirable property for any well-formed typology or classification, as the aim is to offer clear distinctions between concepts [130]. Strong classificatory principles and mutual exclusiveness were therefore demanded for the study at hand. The classification in [7] is also very general in nature, yielding rather abstract anomaly types. This is made clear by the fact that Type I to VI occurrences can all manifest themselves as a point anomaly, and a similar argument holds for contextual anomalies. In order to provide a concrete understanding, the current study offers not only a high-level framework hosting 3 broad groups and 9 basic types, but also a full typology with 63 tangible and detailed subtypes.

**Conceptual variations.** The concept of the anomaly refers to occurrences being both rare and different [9, 17, 319]. This can be seen as ultimately referring to anomalies being non-concentrated and positioned in the lowest density areas [13, 66, 311], with them being more extreme if they have more attributes that deviate and with values that deviate in a more severe fashion [297, 298]. However, it is worth discussing some nuances, mainly in the context of the typology's data distribution dimension. For example, the global density anomaly (ST-IVd) shows that for a numerical dataset in which random noise is the norm, a small cluster actually forms the anomaly [184, 307]. Similarly, high-density regions may constitute the anomalies in dependent data (ST-VIk-l), such as when detecting geographical hot-spots of traffic accidents or disease outbreaks [66, 277]. Relatedly, categorical data may consist of nearly-unique vales, meaning that values that do repeat form the exception (ST-IIb) [181].
Anomalies can also be deviant on only one or two attributes, and show unremarkable behavior on the other variables [57, 162, 163, 298]. A more complex manifestation is the high-density anomaly (HDA), which is an occurrence that deviates from the norm but in some subspace is located in a high-density region, and as such is positioned amongst or is a member of the most normal cases [297]. In cross-sectional (independent) data, subtypes IIa-b, Va and VIa may turn out to be a HDA in its most basic form, but in principle all 63 subtypes can have high-density counterparts when taking into account an additional subspace. HDAs can be interpreted as deviant occurrences that hide in normality, and are especially relevant in misbehavior detection. Identifying them implies solving a delicate balancing problem, taking both anomalousness and normalness into account. The centrally positioned ST-IIa case in the left panel of Fig. 6 is a HDA while the top one is positioned in a moderate or low density area.

**Domain-specific anomalies.** The AD field features many overviews of anomalies that are specific to a given domain. An example is the classification of [265, cf. 254], which describes a collection of anomaly types in maritime traffic data that may point to erroneous or falsified messages. Examples of types are 'too fast for the given vessel', 'vessel type incompatible with size', 'non-declared flag change' and 'outside of usual roads'. Examples of anomaly types observed on stock markets are 'forward rate bias', 'the new December effect', 'short-term price drift' and 'the weekend effect' [266]. Examples from the domain of computer network traffic are 'port scans', 'denial of service attacks', 'alpha flows' and 'outage events' [267]. Other domain-specific classifications describe land parcel anomalies [268], data center anomalies [269], biological malformations [270], and wireless sensor network threats [152, 271]. Some of these domain types can be detected by unsupervised techniques (and can be typified using this study's typology). However, many types represent deviations from specific patterns and are only relevant for a given domain or problem, and thus require supervised or rule-based detection methods.

# 5 Conclusion

This study has presented a comprehensive theoretical conceptualization of anomalies that offers a concrete understanding of the nature and different types of anomalies in datasets. The contributions of this study can be summarized as follows.

- It presents the first all-encompassing, theoretically principled, data-centric, general and domain-independent typology that offers a tangible understanding of the nature and types of anomalies. Apart from preliminary versions [6, 69, 70] no comparable typology, classification or conceptualization seems to have been published before. An extensive literature review has been conducted to ground the typology and overview of anomaly types in the rich contributions of extant research.
- Rather than presenting a mere summing-up, the anomaly types are discussed in terms of fundamental dimensions, or classificatory principles. These dimensions offer a deep insight into the nature of the theoretical concept of the anomaly, and, as they systematically partition the classificatory space, serve to differentiate between the various mutually exclusive types and subtypes. The study employs five data-oriented dimensions to understand and define the concept of anomalies, and to distinguish between the multitude of anomaly types and subtypes. These five cardinal aspects of anomalies are the data type, cardinality of relationship, anomaly level, data structure and data distribution. By employing these dimensions this work aims to turn the generally "vague" view on anomalies into a grounded and tangible theoretical concept, and to yield a typology that is principled, meaningful, non-arbitrary and offers explanatory power. The typology's framework is presented in Fig. 2 and the full typology, this study's core contribution, is summarized in Fig. 3.

- By using three levels of abstraction the typology offers a hierarchical insight into the different manifestations of anomalies:
    - Three general groups: atomic univariate anomalies, atomic multivariate anomalies, and aggregate anomalies.
    - Nine basic and stable types of anomalies.
    - An extensible set of different subtypes that offer a concrete understanding of how the basic types can manifest themselves in datasets. Based on the data types, cardinality and level, the subtypes can be positioned principally and logically in one of the framework's nine main anomaly type cells, within which they can be described in more detail using the data structure and distribution. This study has presented 63 subtypes, but future research may discover new ones, for example as a result of entirely new data structures. Fig. 3 provides a summary of all subtypes.
- The typology can be used to meaningfully comprehend and explain the results of data analyses in both academia and practice, and to evaluate anomaly detection algorithms in a transparent and understandable fashion. It facilitates the creation of test sets and serves to clarify which types and subtypes can and cannot be detected by different (versions or parameterizations of) AD algorithms.
- More in general, this study shows that attention should not merely be paid to algorithms and the detection process, but also to understanding the anomalies themselves by using a data-centric perspective. This helps to comprehend and explain the data and ultimately the world, and may offer opportunities for developing new knowledge. For this reason both academics and practitioners would do well to not see the field as limited to anomaly detection, but to adopt the broader perspective of anomaly analysis and detection (AAAD). After all, in addition to detecting anomalies it is important to understand and explain why a given occurrence is anomalous, especially because follow-up actions are often required to manage the identified deviations.
- Finally, the typology can be employed to scope and position academic studies, and to structure publications, courses, lectures, tutorials and projects on the basis of, for example, three general anomaly groups or nine basic types.

With these contributions this study aims to advance the maturity of the field. Not only by presenting a comprehensive overview of anomaly types, but also by offering an overarching, integrative and fundamental conceptualization of the field's focal topic, the anomaly. This research has centered its attention on anomaly types that can be meaningfully described in terms of data and can be identified by unsupervised AD methods. Future research may therefore extend the scope to rare classes and categories that require (semi-)supervised methods to be detected [164, 272]. Research has shown that these share various characteristics with anomalous occurrences that may be unlabeled and detectable by unsupervised methods [182, 196]. As the current study has primarily focused on anomalies that can be characterized by their intrinsic data properties and thus be detected in an unsupervised mode, this similarity may bring with it interesting opportunities. Another topic for further research is studying in more detail how the data distribution can be used for further classification and clarification of anomaly subtypes. The way in which the distribution impacts anomalies in graphs and other complex data structures can also be studied, for example in the context of the current trend of AD in data streams. Regardless of future work, it is hoped this study has shown the inspiring richness of anomaly analysis and detection as well as the many contributions this field can make to both science and practice.

**Remarks.** This research has been partly supported by HEINEKEN, Loonaangifteketen and UWV, but did not receive a specific grant from funding agencies, nor is there any conflict of interest. The author thanks Arno Schilperoord and the anonymous reviewers for their valuable contributions. An earlier version of this article was published on arXiv on July 30, 2020. Preliminary versions of the typology were presented in [69] and the poster of [6]. Feel free to contact the author if anomaly (sub)types are deemed missing.

Also see the video *Taming the Anomaly: An Overview of Outliers and Other Deviations in Data*, which uses animated 3D data plots to illustrate the various anomaly (sub)types: https://youtu.be/XOBR2sc00y4